\documentclass[aps,prx,reprint,floatfix,superscriptaddress,footinbib,twocolumn]{revtex4}
\usepackage{xcolor}
\usepackage{microtype}
\usepackage[bookmarks=true,
bookmarksnumbered=false, 
bookmarksopen=false, 
colorlinks=true,
linkcolor=blue]{hyperref}
\definecolor{webblue}{rgb}{0, 0, 0.5} 
\hypersetup{colorlinks=true,urlcolor=blue,citecolor=blue}
\usepackage{graphicx}
\usepackage{subfigure}
\usepackage{url}
\usepackage{hyperref}
\usepackage{inconsolata}
\usepackage{amsmath}
\usepackage[capitalize]{cleveref}
\usepackage{braket}
\usepackage{outlines}
\usepackage{enumitem}
\usepackage{soul}
\usepackage{color}

\usepackage{bm} 
\DeclareMathOperator{\Tr}{Tr}
\usepackage[utf8]{inputenc}
\usepackage{siunitx,booktabs}
\usepackage{multirow}
\usepackage{amssymb}

\usepackage{relsize}

\usepackage{soul}
\usepackage{tabularx}

\begin{document}

 \title{Higher-order topological superconductivity from repulsive interactions in kagome and honeycomb systems}
\author{Tommy Li}
\affiliation{Dahlem Center for Complex Quantum Systems and Fachbereich Physik, Freie Universit\"{a}t Berlin, Arnimallee 14, 14195 Berlin, Germany}
\author{Max Geier}
\affiliation{Dahlem Center for Complex Quantum Systems and Fachbereich Physik, Freie Universit\"{a}t Berlin, Arnimallee 14, 14195 Berlin, Germany}
\affiliation{Center for Quantum Devices, Niels Bohr Institute, University of Copenhagen, DK-2100 Copenhagen, Denmark}
\author{Julian Ingham}
\affiliation{Physics Department, Boston University, Commonwealth Avenue, Boston, MA 02215, USA}
\author{Harley D. Scammell}
\affiliation{School of Physics, University of New South Wales, Sydney 2052, Australia}
\affiliation{Australian Research Council Centre of Excellence in Future Low-Energy Electronics Technologies, University of New South Wales, Sydney 2052, Australia}


\begin{abstract}
{ We discuss a pairing mechanism in interacting two-dimensional multipartite lattices that intrinsically leads to a second order topological superconducting state with a spatially modulated gap. When the chemical potential is close to Dirac points, oppositely moving electrons on the Fermi surface undergo an interference phenomenon in which the Berry phase converts a repulsive electron-electron interaction into an effective attraction. The topology of the superconducting phase manifests as gapped edge modes in the quasiparticle spectrum and Majorana Kramers pairs at the corners. We present symmetry arguments which constrain the possible form of the electron-electron interactions in these systems and classify the possible superconducting phases which result. Exact diagonalization of the Bogoliubov-de Gennes Hamiltonian confirms the existence of gapped edge states and Majorana corner states, which strongly depend on the spatial structure of the gap. Possible applications to vanadium-based superconducting kagome metals AV$_3$Sb$_3$ (A=K,Rb,Cs) are discussed.
}
\end{abstract}

\maketitle

\section{Introduction}
Since the discovery of high-temperature superconductors, investigations of superconducting instabilities driven purely by the repulsive Coulomb interaction have led to the discovery of numerous unconventional superconducting phases \cite{Sigrist2005,Nest,Leggett1975,Laughlin1998,Senthil1999,Horovitz2003,Mackenzie2003,Kohn1965}. Among these, topological superconductors -- hosting robust, zero-energy edge modes -- have taken a key role in the pursuit of platforms for quantum computing \cite{Volovik1997, ReadGreen2000,Nandkishore2012}. Recently, a new classification of topological superconductors has emerged which relies on crystalline symmetries in addition to non-spatial ones \cite{Peng2017,Langbehn2017,Geier2018,Geier2020,Trifunovic2019,Trifunovic2020,Shiozaki2019,Ono2020,Khalaf2018,Zhang2020,Gray2019,Choi2020}. These “higher-order” topological superconductors host non-trivial boundary states with multiple dimensionalities: in two dimensions, the second-order topological superconducting phase hosts a gapped spectrum of propagating modes along the edge as well as zero-energy modes localized at the corners of the system in addition to a gapped bulk quasiparticle spectrum.

In the search for candidate materials hosting these phases, it is crucial to investigate the possible mechanisms that might give rise to them in physical systems, based on a microscopic analysis of the electron-electron interactions. In this work we describe an approach to interactions in two dimensional materials with Dirac points that naturally leads to a second order topological $p+i\tau p$ superconducting phase, arising past a critical doping for arbitrarily weak values of the repulsive Coulomb interaction. Our analysis is performed in the weak coupling regime, in which Stoner instabilities and correlated insulating states are absent, but the effects we describe dominate in materials where the atomic orbitals are strongly localized, which distinguishes our theory from those previously used to study graphene.

The Coulomb repulsion is converted into an attraction via destructive interference for certain scattering channels near the Dirac points which originate from the Berry phase, causing the formation of Cooper pairs to lower the correlation energy. Thus, our theory relies on Dirac Fermi surfaces, a scenario distinct from tuning the Fermi level to a nesting density or van Hove singularity, commonly studied as an explanation for superconductivity \cite{Labbe1987,Dzyaloshinskii1987,Friedel1989,Markiewicz1997,Maiti2010,McChesney2010,Kiesel2012}. In contrast to previously studied mechanisms for higher order topological superconductivity, we also do not require proximitization to an existing superconductor \cite{Zhang2019,Zhu2019,Franca2019,Wu2019,Roy2020,Anh2020,Zhang2021,Chew2021}, negative-$U$ Hubbard interactions \cite{Hsu2020}, or bosonic fluctuations \cite{Wang2018}.

This pairing mechanism was previously studied in the context of an artificial honeycomb lattice \cite{Li2020}; here we classify the resulting superconducting states and demonstrate their topological properties. Our results follow from general symmetry considerations, but we will present specific results for kagome and honeycomb lattices. In Section \ref{Sec2} we discuss Hubbard-like models of these systems, presenting the most general set of interactions consistent with the lattice symmetries and enumerating the possible superconducting states which result. In Section \ref{Sec3} we discuss how the superconducting instabilities are affected by screening effects. In Section \ref{Sec4} we perform exact diagonalization of the Bogoliubov-de Gennes Hamiltonian describing the superconducting states, confirming the existence of edge and corner modes for samples whose edge respects the point group symmetry of the lattice. We conclude with a discussion of candidate materials, and suggest the mechanism can explain recently observed superconductivity in vanadium-based kagome metals AV$_3$Sb$_3$ (A=K,Rb,Cs), where superconductivity appears to be unconventional but correlations are weak \cite{Ortiz2020,Zhu2021,Chen2021,Ortiz2021,Ni2021,Chenb2021,Liang2021}.

\section{Superconducting instabilities in Dirac materials}
\label{Sec2}

\begin{table*}
\begin{tabularx}{\textwidth}{lllll}
\hline \hline \\[-2.2mm]
\hspace{6mm} & \ \ \ \  \ \ \ \  & Kagome & \ \ \ \  \ \ \ \  & Honeycomb \\ \\[-2.9mm] \hline \\[-2.4mm]
$g_1/\Omega$ \hspace{6mm}& \ \ \ \  \ \ \ \  & $\frac{1}{6}\left(2U_{AAAA} + 8U_{ABAB}+ 8U_{AABA} + 4U_{ACBC} + U_{AABB}\right)$ & \ \ \ \  \ \ \ \  & $\frac{1}{2}\left(U_{AAAA} + 3 U_{ABAB}\right)$ \\[2mm]
$g_2/\Omega$ \hspace{6mm}& \ \ \ \  \ \ \ \  & $\frac{1}{2} U_{AABB}$ & \ \ \ \  \ \ \ \  &$\frac{1}{2}\left(U_{AAAA} - 3 U_{ABAB}\right)$ \\[2mm]
$g_3/\Omega$ \hspace{6mm}& \ \ \ \  \ \ \ \  & $\frac{1}{3}\left( U_{AAAA} - 2U_{ABAB} + 4U_{AABA} - 4U_{ACBC} + 2U_{AABB}\right)$ & \ \ \ \  \ \ \ \  & $3 U_{ABBA}$ \\[2mm]
$g_4/\Omega$ \hspace{6mm}& \ \ \ \  \ \ \ \  & $\frac{1}{3}\left( U_{AAAA} - 2U_{ABAB} + 3U_{AABA} - 3U_{ACBC} + 2U_{AABB} \right)$ & \ \ \ \  \ \ \ \  & $3 U_{ABBA}$ \\[2mm]
$g_5/\Omega$ \hspace{6mm}& \ \ \ \  \ \ \ \  & $\frac{1}{3}\left( U_{AAAA} + U_{ABAB} + 4U_{AABA} + 8U_{ACBC} + 8U_{AABB} \right)$ & \ \ \ \  \ \ \ \  & $U_{AAAA}$ \\[2mm]
\hline \hline
\end{tabularx} 
\caption{The relations between coupling constants (\ref{Hamiltonian_density}) and the extended Hubbard model parameters (\ref{Hamiltonian:lattice}) for the Kagome and honeycomb lattice with couplings involving nearest neighbors. We denote $U_{\sigma_1\sigma_2\sigma_3\sigma_4} = U(\bm{r}_1,\bm{r}_2,\bm{r}_3,\bm{r}_4)$ where $\bm{r}_1,\bm{r}_2,\bm{r}_3,\bm{r}_4$ are sites in sublattices $\sigma_1,\sigma_2,\sigma_3,\sigma_4$  separated at most by a nearest neighbor bond.}
\label{tab:couplings}
\end{table*}
We investigate the pairing instabilities of a two-dimensional lattice with hexagonal symmetry hosting Dirac points at the $K$ points, which we describe via an extended Hubbard model
\begin{gather}
H = \sum_{\bm{r},\bm{r}'}{T(\bm{r}-\bm{r}') c^\dag_{\bm{r}} c_{\bm{r}'}} \nonumber \\
+ \frac{1}{2}\sum_{\bm{r}_1,\bm{r}_2,\bm{r}_3,\bm{r}_4}{
U(\bm{r}_1,\bm{r}_2,\bm{r}_3,\bm{r}_4)(c^\dag_{\bm{r}_3} c_{\bm{r}_1})
(c^\dag_{\bm{r}_4} c_{\bm{r}_2})}
\label{Hamiltonian:lattice}
\end{gather}
where $c^\dag_{\bm{r}} = (c^\dag_{\bm{r},\uparrow},c^\dag_{\bm{r},\downarrow})$ are two component spinors and the spin inner product is implied in the single particle and interacting terms in the Hamiltonian (\ref{Hamiltonian:lattice}).

We choose lattice vectors $\bm{a}_1 = (a,0)$, $\bm{a}_2 = (\frac{a}{2},\frac{\sqrt{3}a}{2})$, and specialize to lattices possessing $C_{6v}$ symmetry, which consists of twofold and threefold rotations in the plane and mirror reflections about the $x$ and $y$ axes. We only consider models without spin-orbit interaction, thus we may choose a basis of Wannier orbitals so that $T(\bm{r}-\bm{r}')$ and $U(\bm{r}_1,\bm{r}_2,\bm{r}_3,\bm{r}_4)$ are real. The Hamiltonian is invariant under SU(2) spin rotations and possesses a time reversal symmetry satisfying $\mathcal{T}^2 = 1$, which is identical to complex conjugation in the coordinate representation. For the remainder of this paper we denote the time reversal operation which leaves spin unaffected by $\mathcal{T}$.

{The resulting band structure features Dirac points at the $K$ and $K'$ points, which we distinguish by a valley index $\tau = \pm $. The eigenstates} of the single particle Hamiltonian may be classified by their eigenvalues $e^{\frac{2\pi i}{3} \alpha}$  under threefold rotations, where $\alpha = \pm $. We shall refer to the $\alpha$ degree of freedom as \emph{pseudospin}. At the Dirac points there are four degenerate energy eigenstates $|\tau,\alpha\rangle$ which transform into one another under time reversal, twofold rotations and mirror symmetry. We shall introduce the Pauli operators $\tau_\mu,\alpha_\mu$ acting on the valley and pseudospin degrees of freedom. The degeneracy of the $\alpha$ eigenstates at each valley is protected by either $M_x$ or $R_{\pi}\mathcal{T}$. We obtain the following representation of crystal symmetries:
\begin{gather}
U_{\mathcal{R}_{\frac{2\pi}{3}}} = e^{\frac{2\pi i}{3} \alpha_z}  , \ \ U_{M_x} = \alpha_x  , \nonumber \\
U_{\mathcal{R}_\pi} = \cos \phi_I \tau_x + \sin \phi_I \tau_y  , \ \ \mathcal{T} = \tau_x\alpha_x \mathcal{K} 
\label{rep:symmetries}
\end{gather}
where $\phi_I$ is a phase which generally depends on the lattice and the chosen center of inversion. 

Doping slightly above the valleys $\pm\bm{K} = \pm (\frac{4\pi}{3a},0)$, the Fermi surface consists of two circular pockets surrounding the $K$ points with Fermi momenta $k_F \ll |\bm{K}|$, with the single particle energy eigenstates formed from linear combinations of the pseudospin basis states (i.e. eigenstates of threefold rotations). By obtaining the band structure of \eqref{Hamiltonian:lattice} and projecting the interactions onto states near the Dirac points, we obtain a quantum field theory describing interaction processes close to the Fermi level which accounts for the $\tau$, $\alpha$ and spin ($s$) degrees of freedom and may be formulated using an 8-component local field operator $\psi^\dag_{\tau\alpha s}(\bm{r})$. The most general Hamiltonian density consistent with the symmetries \eqref{rep:symmetries} is of the form \begin{gather}
\mathcal{H} = \psi^\dag (-iv \tau_z\bm{\alpha}\cdot\nabla)\psi +\tfrac{1}{2} \sum_{a,b}{\mathcal{V}_{ab} \left(\psi^\dag J^a \psi\right)\left(\psi^\dag J^b\psi\right)}
\label{Hamiltonian_density}
\end{gather}
where we have expanded the interaction in the adjoint basis consisting of products  $J^a = \tau_\mu \alpha_\nu$, and $\mathcal{V}_{ab}$ are constants which may be obtained by projecting the extended Hubbard interactions (\ref{Hamiltonian:lattice}) onto the pseudospin and valley eigenstates. As we show in the Appendix, only operators which are even under time reversal symmetry are permitted in the interactions, which are $\{J^1, J^2,\dots, J^{10}\} = \{\tau_0\alpha_0, \tau_z\alpha_z, \tau_0\alpha_x, \tau_0\alpha_y, \tau_x, \tau_y, \tau_x\alpha_x, \tau_x\alpha_y, \tau_y\alpha_x, \tau_y\alpha_y\}$. The Hamiltonian must transform as a scalar under the spatial symmetries (\ref{rep:symmetries}) which implies that $\mathcal{V}$ is a diagonal matrix. There are only five independent coupling constants $g_1,g_2,g_3,g_4,g_5$, which we relate to the interaction matrix $\mathcal{V}_{ab}$ by
\begin{gather}
\mathcal{V}_{11} = g_1  , \  \mathcal{V}_{22} = g_2   , \  \mathcal{V}_{33} =\mathcal{V}_{44}= \frac{g_3}{2} , \nonumber \\
\mathcal{V}_{55} = \mathcal{V}_{66} = \frac{g_4}{2} \  ,\  \mathcal{V}_{77} = \mathcal{V}_{88} = \mathcal{V}_{99} = \mathcal{V}_{10,10} = \frac{g_5}{4}  \ .
\label{V_g}
\end{gather}
In addition to a density-density interaction $g_1$, the interactions involve valley-conserving ($J^2 =\tau_z\alpha_z$) and valley-mixing ($J^5 = \tau_x, J^6 = \tau_y$) mass operators. The remaining six operators are conserved SU(2) valley currents which can be expressed as products of the valley operators $i\tau_x,i\tau_y,\tau_z$ and the electric current $j_\mu = (j_x,j_y) = (\tau_z\alpha_x,\tau_z\alpha_y)$.

The relation between the coupling constants $g_i$ appearing in the field theory (\ref{Hamiltonian_density}) and the extended Hubbard model parameters $U(\bm{r}_1,\bm{r}_2,\bm{r}_3,\bm{r}_4)$ for the kagome and honeycomb lattices are given in Table ~\ref{tab:couplings}. We denote the three sublattices of the kagome lattice $\{A,B,C\}$, and the two sublattices of the honeycomb lattice by $\{A,B\}$. We list only interactions involving nearest neighbors, and denote the interactions in which the lattice coordinates $\bm{r}_1,\bm{r}_2,\bm{r}_3,\bm{r}_4$ exist on the $\sigma_1,\sigma_2,\sigma_3,\sigma_4$ sublattices by $U_{\sigma_1\sigma_2\sigma_3\sigma_4}$. In the kagome lattice, each site has two nearest neighbors in each of the other sublattices, and in the honeycomb lattice, each site has three nearest neighbors in the other sublattice.

The eigenstates of the Dirac Hamiltonian in the upper band are given by $f^\dag_{\bm{k},\tau,s} = (\psi^\dag_{\tau \tau s}(\bm{k}) + \tau e^{i\tau \theta_{\bm{k}}} \psi^\dag_{\tau -\tau s}(\bm{k}))/\sqrt{2}$. Evaluating the Born amplitudes from (\ref{Hamiltonian_density}) we find that the scattering vertex $\Gamma_{\tau_1\tau_2\tau_3\tau_4}(\theta)$ in the Cooper channel is given by
\begin{align}
&\Gamma_{\tau\tau\tau\tau}(\theta) = \frac{g_1+g_2}{2} e^{-i\tau \theta}\cos\theta  + \frac{g_1-g_2-g_3}{2} e^{-i\tau\theta} \ \ , \nonumber \\
&\Gamma_{+-+-}(\theta) = \Gamma_{-+-+}(\theta) = \frac{g_1 -g_2}{2}\cos\theta + \frac{g_1+g_2+g_3}{2} \ \ , \nonumber \\
&\Gamma_{+--+}(\theta) = \Gamma_{-++-}(\theta) = -\frac{g_4}{2}\cos\theta + \frac{g_4+g_5}{2} \ \ .
\label{BCS:vertex}
\end{align}
where $\theta$ is the scattering angle. BCS theory predicts that superconducting pairing occurs when the scattering amplitude between Cooper pairs is negative, i.e. for attractive scattering. For pairing of electrons within the same valley $\tau$ this corresponds to $\Gamma^\ell_{\tau\tau\tau\tau} < 0$, while pairing of electrons in opposite valleys requires the symmetrized amplitudes $\Gamma^\ell_{+-+-} + \Gamma^\ell_{+--+} < 0$ or $\Gamma^\ell_{+-+-} - \Gamma^\ell_{-++-} < 0$. The critical temperature is given 
\begin{gather}
\label{Tc}
T_c \sim E_F e^{-1/(\nu_0\lambda)}
\end{gather}
where $\lambda = \Gamma^\ell_{\tau\tau\tau\tau}$ for intravalley pairing or $\lambda = \Gamma^{\ell}_{+-+-} \pm \Gamma^{\ell}_{+--+}$ for intervalley pairing, and $\nu_0 = k_F/(2\pi v)$ is the density of states per spin per valley at the Fermi level.

In order to investigate the physical features of the condensate we introduce the mean field Hamiltonian
\begin{align}
H_\text{BdG}&=\sum_{\bm{k}}\varepsilon_{\bm k} f^\dag_{\bm k}f_{\bm k} + \frac{1}{2}f^\dag_{\bm k} \left(\Delta_{\bm k} i s_y i \tau_y\right) f^\dagger_{-\bm k} + \text{h.c.} 
\end{align}
where $f^\dag_{\bm{k}}$ is the creation operator for a Dirac fermion in the upper band and $\Delta(\bm{k})$ is the superconducting gap matrix (spin and valley indices are implied). Accounting for spin, valley and angular momentum structure, there are eight possible superconducting phases, which we list in Table ~\ref{tab:gaps} along with the corresponding scattering amplitude.

The topological properties of the gaps are associated with their Altland-Zirnbauer class \cite{Altland1996,Chiu2016,Kitaev2009}, which classifies mean field Hamiltonians based on time reversal, charge conjugation and chiral symmetries. For the spin singlet pairing phases, $\Delta \propto s^0$, accounting for SU(2) spin rotation symmetry we find that the charge conjugation symmetry satisfies $\mathcal{C}^2 = -1$. In the case of spin triplet pairing, the spontaneous violation of SU(2) spin rotation symmetry fixes a direction $\bm{d} = (d^x, d^y, d^z)$, with the gap $\Delta \propto d^\mu s_\mu $. As we show in the Appendix, magnetization of the condensate is energetically disfavored, which implies that $\bm{d}$ can be chosen to be purely real. The mean-field Hamiltonian is invariant under spin rotation about the  $\bm{d}$-direction. Accounting for this U(1) symmetry \footnote{We note that the pairing term only possesses U(1) spin symmetry, however the normal state dispersion retains its original SU(2) spin symmetry. It was noted in \cite{Schnyder2008} that such a state generically falls in class D.}, we may decompose the mean field Hamiltonian in two spin blocks, with each separately possessing a charge conjugation symmetry satisfying $\mathcal{C}^2 = +1$. We therefore conclude that the phases with spin singlet pairing are either in class CI or C, while those with spin triplet pairing are either in class BDI or D depending on whether the time-reversal symmetry of the normal state Hamiltonian survives in the condensed phase. Consulting the periodic table of topological invariants \cite{Kitaev2009}, we find that first order topological superconductivity is possible for the phases in which time reversal symmetry is spontaneously broken, i.e. those in class C or D, and posses a Chern number which counts the number of chiral modes propagating along the boundary and whose parity is equal to the parity of Majorana modes localized at a vortex. However, both the intravalley $p$-wave spin triplet, and intervalley $s$-wave spin triplet states are in class BDI. For the intravalley spin triplet phase, the gap is $p+ip$ in one valley and $p-ip$ in the other, preserving time reversal symmetry, and we refer to this as the $p+i\tau p$ phase. In these cases lowest order topology is absent while second order topology is possible \cite{Geier2018, Trifunovic2019,Khalaf2018}. These phases are indicated by the existence of Majorana corner states of definite spin protected by crystalline symmetries. In addition to the spinless time reversal symmetry operator satisfying $\mathcal{T}^2 = +1$, our system possesses a spinful time reversal symmetry $\mathcal{T}^\prime = \mathcal{T} e^{i\pi S_y}$ satisfying $(\mathcal{T}^\prime)^2 = -1$ which interchanges opposite spin blocks; thus corner modes occur in Kramers pairs. In Section \ref{Sec4} we present exact diagonalization results which demonstrate that the intravalley $p+i\tau p$ spin triplet phase indeed possesses a second order topology.

\begin{table*}
\begin{tabularx}{\textwidth}{llllllc}
\hline \hline \\[-2.2mm]
Gap structure & \ \ \ \ \ & $\Delta$ & \ \ \ \ \ &  $\lambda $ & \ \ \ \ \ & AZ class \\ \\[-2.9mm] \hline \\[-2.4mm]
Intervalley, spin singlet, $s$-wave & \ \ \ \ \ & $  \tau_z$ & \ \ \ \ \ & $\frac{1}{2}\left(g_1+g_2+g_3+g_4+g_5\right)$ & \ \ \ \ \ & CI \\
Intervalley, spin singlet, $p$-wave & \ \ \ \ \ & $  e^{\pm i \theta_{\bm{k}}} $ & \ \ \ \ \ & $\frac{1}{4}\left(g_1-g_2+g_4\right)$ & \ \ \ \ \ & C \\
Intervalley, spin triplet, $s$-wave & \ \ \ \ \ & $ d^\mu s_\mu $ & \ \ \ \ \ &  $
\frac{1}{2}\left(g_1+g_2+g_3-g_4-g_5\right)$ & \ \ \ \ \ & BDI \\
Intervalley, spin triplet, $p$-wave & \ \ \ \ \ & $  e^{\pm i \theta_{\bm{k}}} \tau_z (d^\mu s_\mu)$ & \ \ \ \ \ & $\frac{1}{4}\left(g_1 - g_2 - g_4\right)$ & \ \ \ \ \ & D \\
Intravalley, spin singlet, $s$-wave & \ \ \ \ \ & $  e^{i\tau_z\phi} (i\tau_y)$ & \ \ \ \ \ & $ \frac{1}{4}\left(g_1+g_2\right)$ & \ \ \ \ \ &  CI \\
Intravalley, spin singlet, $d$-wave & \ \ \ \ \ & $  e^{i\tau_z(\phi - 2\theta_{\bm{k}})} (i\tau_y)$ & \ \ \ \ \ & $\frac{1}{4}\left(g_1+g_2\right)$ & \ \ \ \ \ & CI \\
Intravalley, spin singlet, mixed $s$/$d$-wave & \ \ \ \ \ & $ e^{\pm i\theta_{\bm{k}}}
 e^{i\tau_z(\phi+\theta_{\bm{k}})}
(i\tau_y)$ & \ \ \ \ \ & $\frac{1}{4}\left(g_1+g_2\right)$ & \ \ \ \ \ & C \\
Intravalley, spin triplet, $p$-wave & \ \ \ \ \ & $  e^{i\tau_z(\phi-\theta_{\bm{k}})} (d^\mu s_\mu)(i\tau_y)$ & \ \ \ \ \ & $\frac{1}{2}\left(g_1-g_2-g_3\right)$ & \ \ \ \ \ & BDI \\[2mm] \hline \hline
\end{tabularx}
\caption{ Superconducting phases for 2D Dirac fermions in lattices with $C_{6v}$ symmetry. \emph{First column}: gap structure, \emph{second column}: order parameter, \emph{third column}: coupling constant which determines $T_c$ via Eq. \eqref{Tc}, in terms of coupling constants in the Hamiltonian Eq. (\ref{Hamiltonian_density}), \emph{fourth column}: Altland-Zirnbauer class.  }
\label{tab:gaps}
\end{table*}

We now apply our previous results to analyze the possibility of superconducting instabilities in the kagome and honeycomb lattices. The leading Hubbard interactions do not involve tunneling between sites, thus we only account for interactions $U(\bm{r}_1,\bm{r}_2,\bm{r}_3,\bm{r}_4)$ in (\ref{Hamiltonian:lattice}) with $\bm{r}_1=\bm{r}_3,\bm{r}_2=\bm{r}_4$. Denoting the parameter $\lambda$ in (\ref{Tc}) by $\lambda_1$ for intervalley $s$-wave spin triplet pairing, $\lambda_2$ for intervalley $p$-wave spin triplet pairing and $\lambda_3$ for intravalley $p$-wave spin triplet pairing, we find for the kagome lattice
\begin{gather}
\lambda_1 =\lambda_2 = \frac{\lambda_3}{2} =  \frac{U_{ABAB}\Omega}{2}
\label{lambda:kagome}
\end{gather}
where $\Omega$ is the area of the unit cell and
\begin{gather}
\lambda_1 = 0 \ \ , \ \lambda_2 = \frac{\lambda_3}{2} =\frac{3 U_{ABAB}\Omega}{4} \ \ .
\label{lambda:honeycomb}
\end{gather}
for the honeycomb lattice.

In the limit $U_{ABAB}\rightarrow 0$ the scattering amplitudes for all three spin triplet instabilities vanish for both lattices, implying that the system approaches a critical point for the onset of triplet superconductivity, including for the topologically nontrivial phases. The vanishing of the scattering amplitude in the $p+ip$ and $p+i\tau p$ channels is a consequence of the Dirac nature of the scattering states, and can be understood as follows: the components of the wavefunction in the pseudospin and valley basis have momentum-dependent phase factors which are a manifestation of the Berry phase surrounding the Dirac points.  Despite our model only containing local interactions, upon projecting the Dirac wavefunction onto the interactions in (\ref{Hamiltonian_density}) we find that the Berry phase gives rise to  $p+ip$ scattering amplitudes which are essential to obtain a nontrivial topology. The pseudospin and valley dependence of the operators $J^a$ allows some $g_i$ to contribute to the partial wave amplitudes $\lambda$ with negative signs, as can be seen in Table  \ref{tab:gaps}. This causes destructive interference between the contributions to the scattering amplitude with distinct pseudospin and valley structure -- i.e. the $g_i$. In the scattering channels which give rise to the topological phases, the contributions from different $g_i$ sum to zero, leading to a total cancellation of the on-site Coulomb repulsion.

An additional negative contribution  to the scattering amplitude which can overcome the weak nearest neighbor repulsion would result in superconductivity. One interesting scenario would be a weak electron-phonon interaction, which in our case could drive the instability to an unconventional spin triplet state, rather than the spin singlet state which normally results. In the next section, however, we will show that even in the absence of additional attractive interactions, a negative contribution to the nearest neighbor repulsion arises  as a result of overscreening. All three triplet amplitudes become negative if the bare nearest neighbor repulsion is sufficiently weak, and the density is sufficiently high.

\section{Screening effects}
\label{Sec3}

We now consider how $g_i$ are modified by screening processes. We show these effects here analytically in the Random Phase Approximation (RPA), equivalent to taking the limit of large $N$ where $N=4$ is the degeneracy of the Dirac points 
\cite{GonzalezN,SonN,Hwang2007,KotovN}. This analysis was performed for the special case of an artificial semiconductor honeycomb lattice in \cite{Li2020}; we extend this analysis to encompass generic Dirac materials satisfying $C_{6v}$ and time-reversal symmetry, which may be described by the Hamiltonian (\ref{Hamiltonian_density}). The RPA results in replacing the interaction constants $\mathcal{V}_{ab}$ in (\ref{Hamiltonian_density}) with screened frequency and momentum dependent couplings, which are determined by the RPA equation 
\begin{align}
    \widetilde{\mathcal{V}}_{ab}(\omega,\bm{q})= \mathcal{V}_{ab} + \sum_{cd}{\mathcal{V}_{ac}\Pi^{cd}(\omega,\bm{q}) \widetilde{\mathcal{V}}_{db}(\omega,\bm{q})},
\end{align}
where $\Pi^{cd}(\omega,\bm{q})$ is a generalized polarization operator. These represent the susceptibility of the system to a perturbation $\delta \mathcal{H} = \sum_a{\phi_a(r,t) J^a}$, and are calculated in the Appendix.

Accounting for the frequency and momentum dependence of the screened scattering vertex, solution of the Eliashberg equations gives the critical temperature 
\begin{gather}
T_c = \Lambda e^{-1/(\nu_0 \widetilde{\lambda})}
\end{gather}
where $\widetilde{\lambda}$ is the RPA renormalised $\lambda$ of Table \ref{tab:gaps} obtained from interactions between electrons on the Fermi surface and $\Lambda$ is a frequency cutoff determined by scattering processes away from the Fermi surface; explicit calculation (discussed in the Appendix) shows $\Lambda \approx E_F$.

Scattering on the Fermi surface corresponds to the range of momentum and frequency transfer $q < 2k_F$, $\omega = 0$. In this range the polarization operators are constant, and $\widetilde{\lambda}$ is simply given by the previous expressions $\lambda$ (Table \ref{tab:gaps}) evaluated with screened couplings $\widetilde{g}_i$ at $\omega=0,q=0$. We have studied the screening of the interactions $g_1$ and $g_2$ in a previous paper \cite{Li2020}, and found that $g_1$ is reduced while $g_2$ is enhanced as the chemical potential is increased. We  can explain the qualitative effects of static and homogeneous screening on the interactions on general physical grounds. The coupling $g_1$ is screened by the long wavelength density-density response, which as usual weakens $g_1$ with increasing chemical potential. The intra and intervalley mass couplings $g_2$ and $g_4$ are screened via the polarization operator associated with the response of the system to an external perturbation $\propto \alpha_z\tau_z, \tau_{x}, \tau_{y}$ which gap the Dirac spectrum. This perturbation lowers the energy of negative energy states and increases the energy of the positive energy states. The response of the occupied negative energy states below the Dirac point is divergent and may be fully subtracted by regularization -- the result of which is that $g_2$ and $g_4$ should be replaced by reduced interaction constants which are different from those provided in Table \ref{tab:couplings}; thus within the Dirac model the only physically significant effect of screening is the increase in energy of the occupied positive energy states which grows stronger with increasing chemical potential. The interactions $g_3,g_5$ are valley current interactions and the polarization operators responsible for their screening also contain a UV divergent part which may be subtracted by replacing the bare interaction constants with reduced values, and after this subtraction are unscreened due to a Ward identity. Explicitly, we find in RPA
\begin{gather}
\widetilde{g}_1 = \frac{g_1}{1 + N\nu_0 g_1} \  , \nonumber\\ \widetilde{g}_i = \frac{g_i}{1 - N\nu_0 g_i} \ , \nonumber\\ \widetilde{g}_j = g_j 
\end{gather}
where $i=2,4$ and $j=3,5$. This results in replacing the nearest-neighbor coupling with a screened value, $U_{ABAB}\rightarrow \widetilde{U}_{ABAB}$ in the relations in Table \ref{tab:couplings} and (\ref{lambda:kagome}), (\ref{lambda:honeycomb}). The screened nearest neighbor interaction is given by
\begin{gather}
\widetilde{U}_{ABAB} = \frac{\widetilde{g}_1-\widetilde{g}_3}{2\Omega} 
= \frac{1}{2\Omega}\left[  \frac{g_1}{1 + N\nu_0 g_1} - g_3\right]
\end{gather}
for the kagome lattice and
\begin{gather}
\widetilde{U}_{ABAB} =  \frac{\widetilde{g}_1-\widetilde{g}_2}{3\Omega} 
=\frac{1}{3\Omega}\left[\frac{g_1}{1 + N\nu_0 g_1} - \frac{g_2}{1 - N \nu_0 g_2}\right]
\end{gather}
for the honeycomb lattice. For sufficiently large $\nu_0$, we find that $\widetilde{U}_{ABAB} <0$, resulting in superconductivity, and the dominant instability occurs in the intravalley, $p$-wave, spin triplet channel. This occurs for $\mu> \mu^*$ with the critical value
\begin{gather}
\mu^* = 2\pi v^2\left[ \frac{g_3^{-1}-g_1^{-1}}{N}\right]
\end{gather}
for the kagome lattice and
\begin{gather}
\mu^* =  2\pi v^2\left[\frac{g_2^{-1} - g_1^{-1}}{2N}\right]
\end{gather}
for the honeycomb lattice.

The Dirac theory is only applicable up to a cutoff beyond which the electronic dispersion becomes nonlinear. Our analysis predicts a superconducting instability when the critical doping $\mu^*$ lies within the regime where Dirac theory is justified. In the limit $g_3\rightarrow 0$ or $g_2\rightarrow 0$ respectively for the kagome and honeycomb lattices, the critical doping becomes infinite. We therefore require significant pseudospin dependent couplings, which are suppressed when the atomic orbitals are delocalized.

So far, we have neglected the long ranged part of the Coulomb interaction, which provides weak couplings between sites with large separation. This may be accounted for by replacing $\widetilde{g}_1$ with its screened value at $q\rightarrow 0$, $\widetilde{g}_1 \rightarrow 1/N\nu_0$, and simply results in a shift of the critical doping to higher values.

\section{Topological edge and corner modes}
\label{Sec4}

\begin{figure*}
\begin{tabular}{lll}
(a) & (b) & (c)\tabularnewline
\hspace{0.12cm} \includegraphics[width=4.5cm]{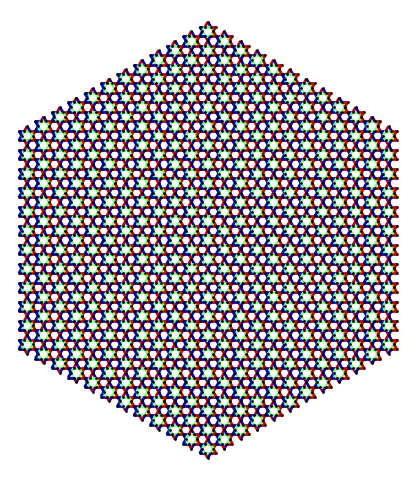} & \hspace{0.15cm} \includegraphics[width=4.5cm]{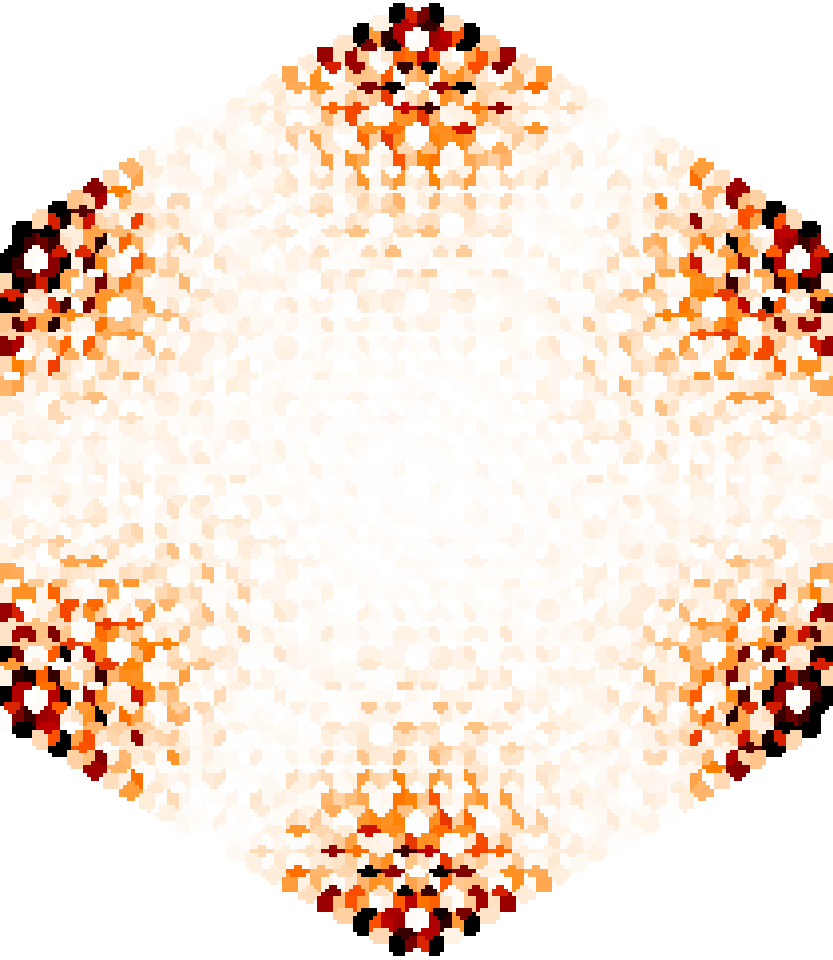} & \includegraphics[width=6cm]{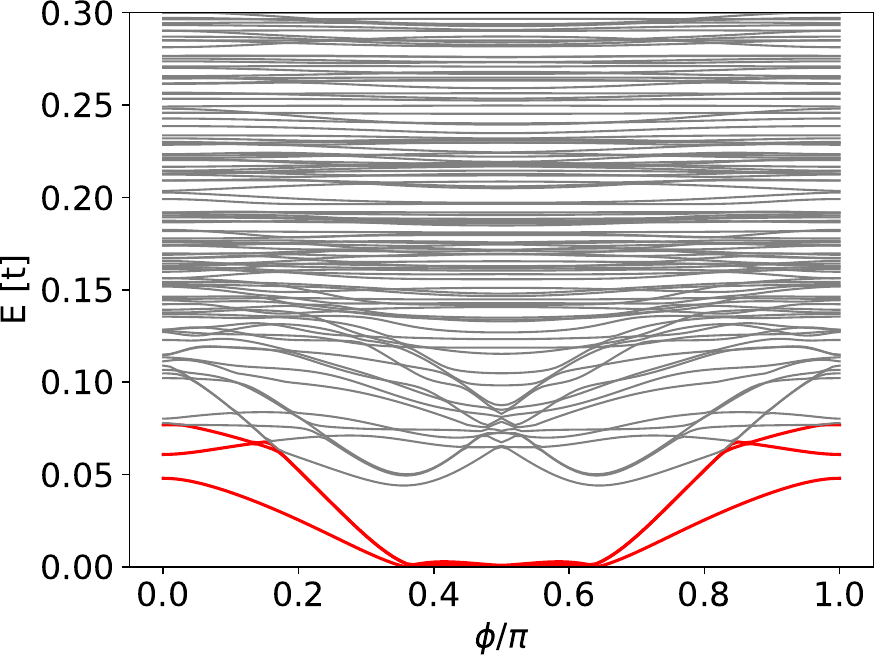}\tabularnewline
(d) & (e) & (f)\tabularnewline
\hspace{0.12cm} \includegraphics[width=4.5cm]{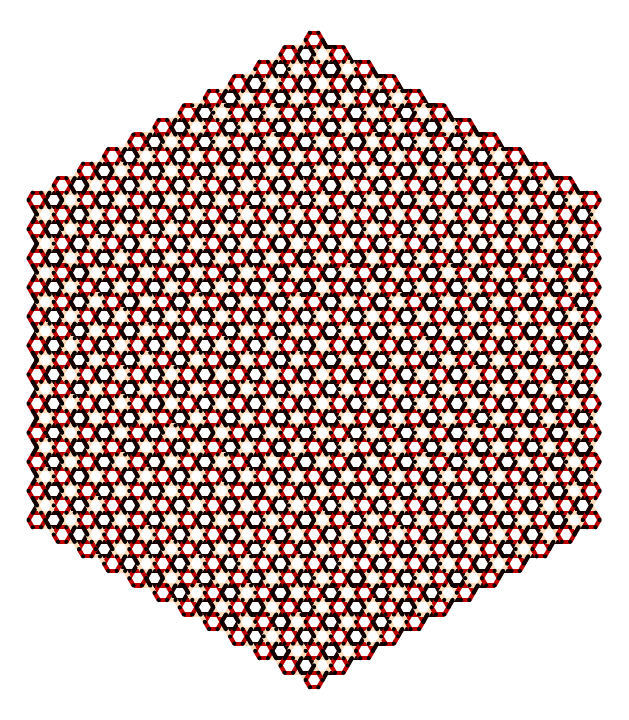} & \hspace{0.15cm} \includegraphics[width=4.5cm]{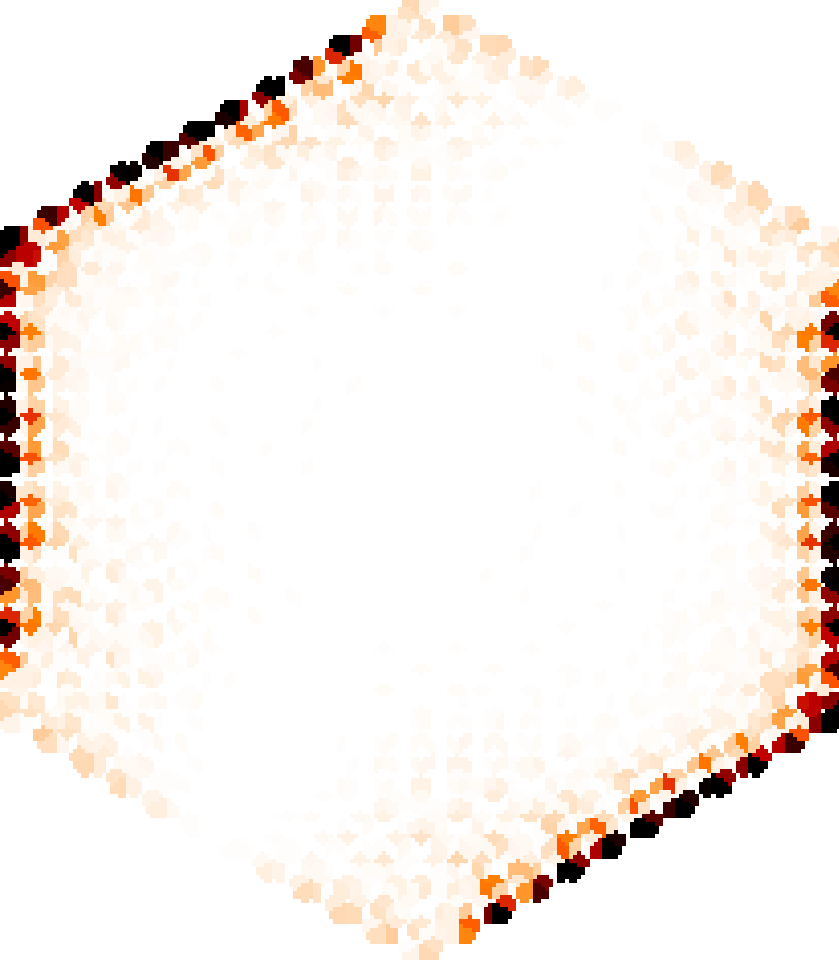} & \includegraphics[width=6cm]{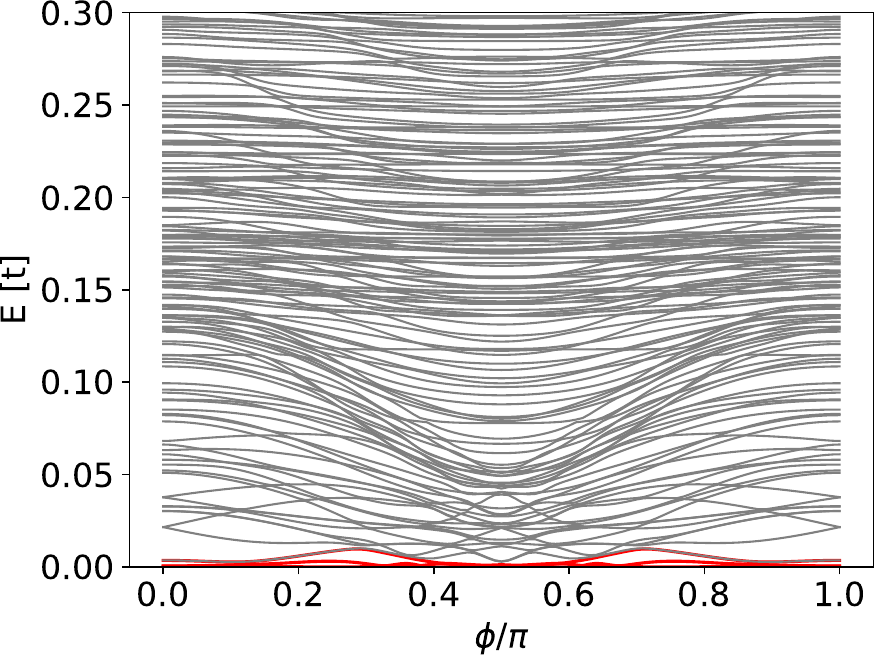}\tabularnewline
(g) & (h) & (i)\tabularnewline
\hspace{-0.3cm}\includegraphics[width=5.5cm]{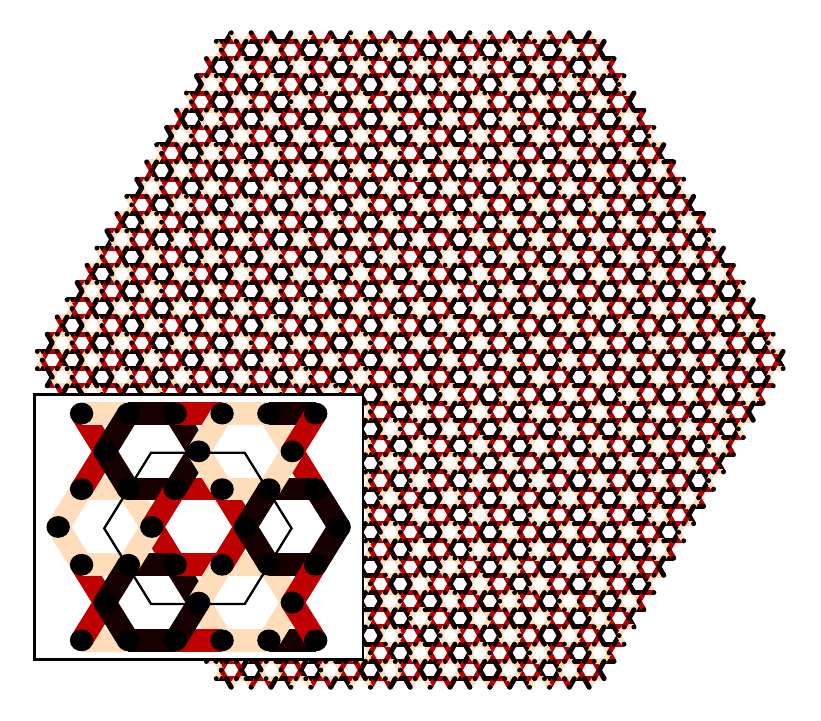} & \hspace{-0.3cm}\includegraphics[width=5.5cm]{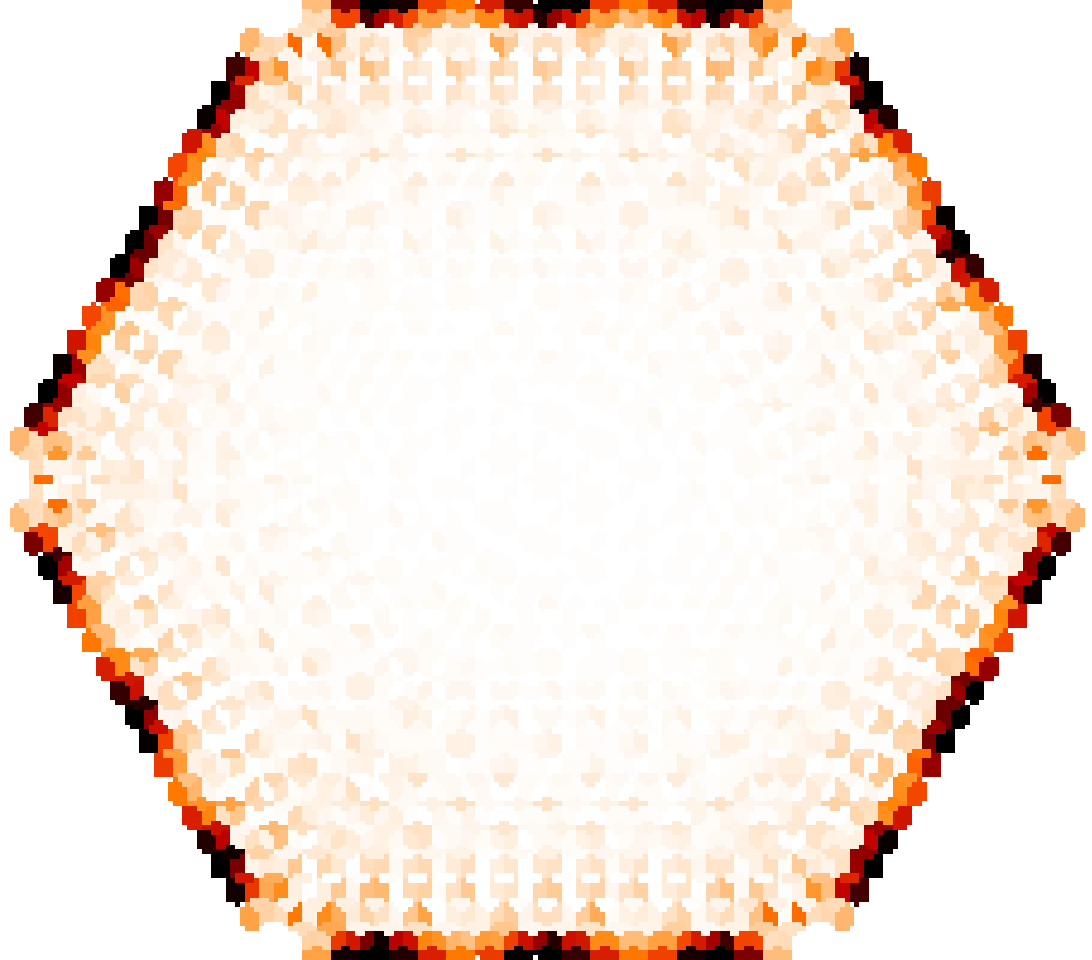} & \hspace{-0.0cm}\includegraphics[width=6cm]{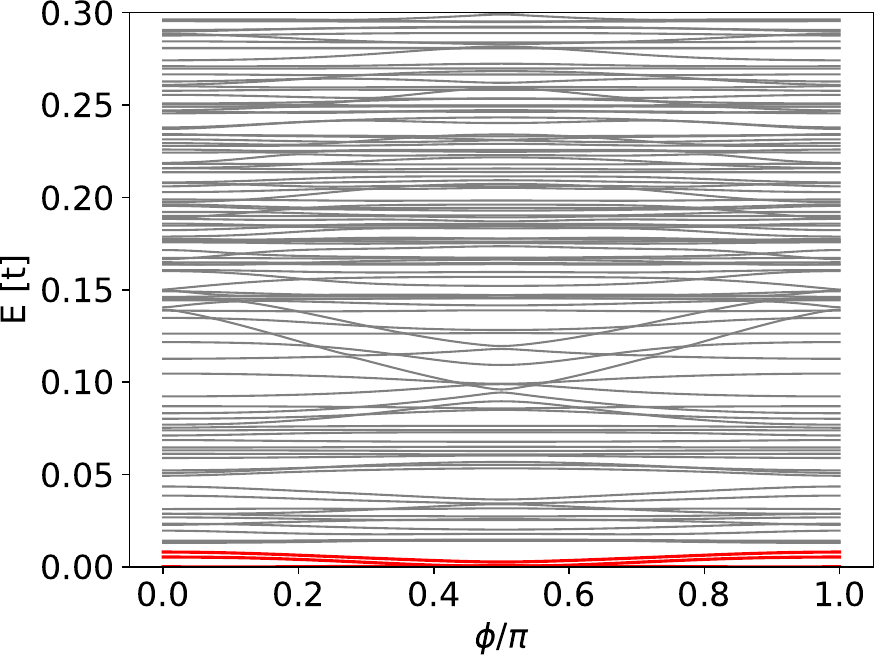}\tabularnewline
\includegraphics[width=5cm]{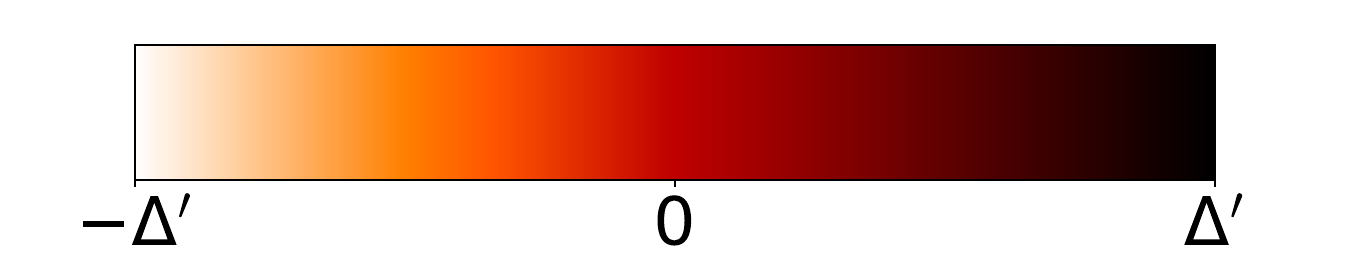} &  & \tabularnewline
\end{tabular}\caption{Results of the exact diagonalization of the Bogoliubov-de Gennes Hamiltonian within a single spin sector for hexagonal flakes of a kagome lattice
of three different geometries. The leftmost figures show the spatially
modulated pairing between nearest neighbors for $\phi=\pi/2$. The
color of the bonds between neighboring sites $\bm{r},\bm{r}'$ shows
the value of the gap function $\Delta(\bm{r},\bm{r}')$ according
to the color scale. The inset of the bottom left figure depicts a
zoom into the bulk of the lattice with the unit cell boundary indicated
by the thin lines. The central figures show the wavefunction densities
of the three lowest quasiparticle modes. The rightmost figures show
the quasiparticle spectrum as a function of the pair density wave
parameter $\phi$. The three lowest quasiparticle modes are indicated
in red. For the exact diagonalization we use the parameters $\mu = 0.4t,\ \Delta^\prime = 0.4t$.}
\label{fig:ED:kagome}
\end{figure*}

\begin{figure*}
\begin{tabular}{lll}
(a) & (b) & (c)\tabularnewline
\hspace{0.15cm} \includegraphics[width=4.5cm]{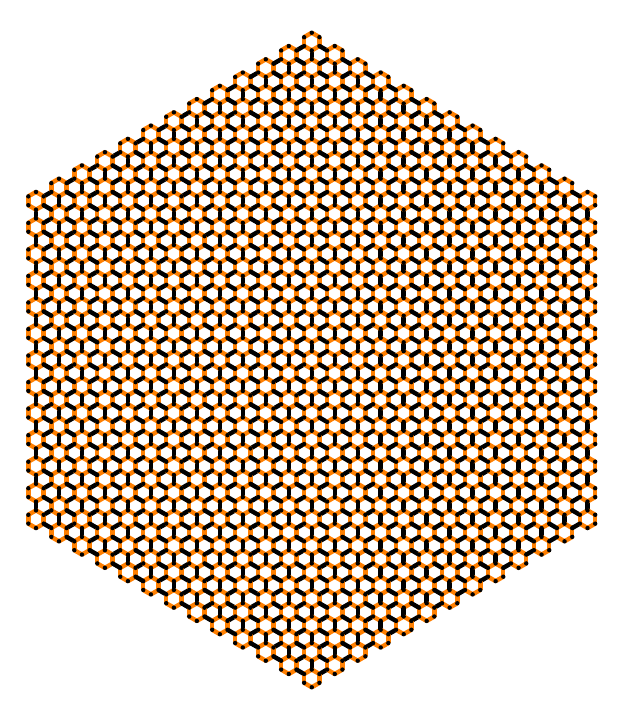} & \hspace{0.15cm} \includegraphics[width=4.5cm]{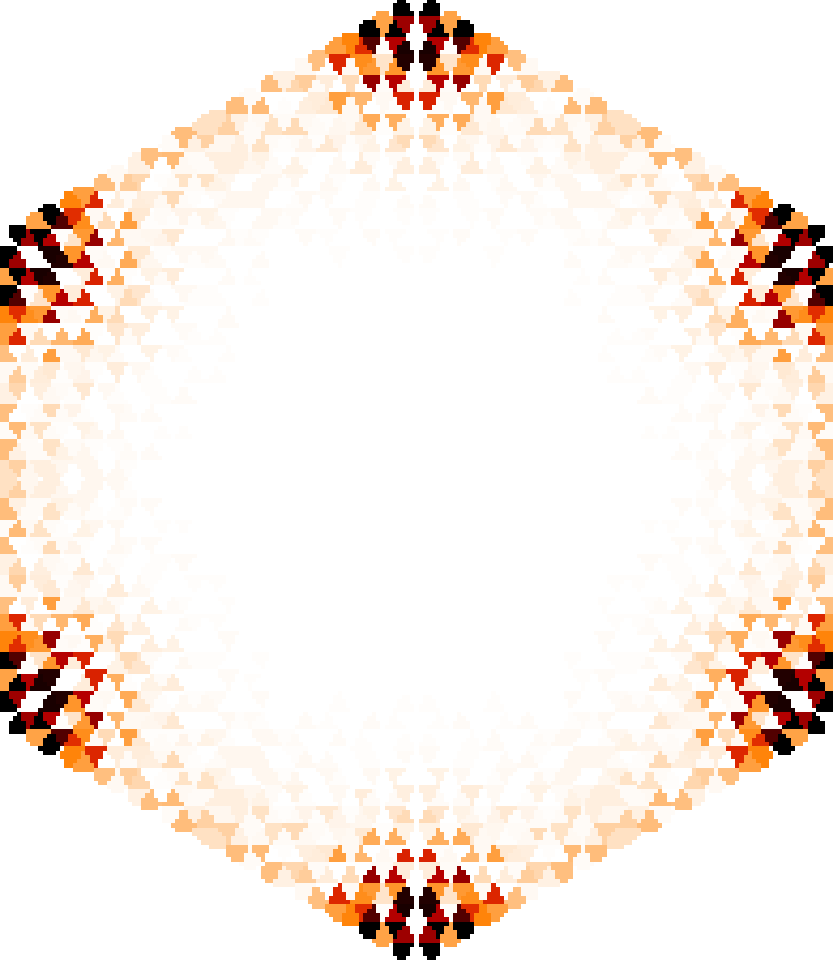} & \includegraphics[width=6cm]{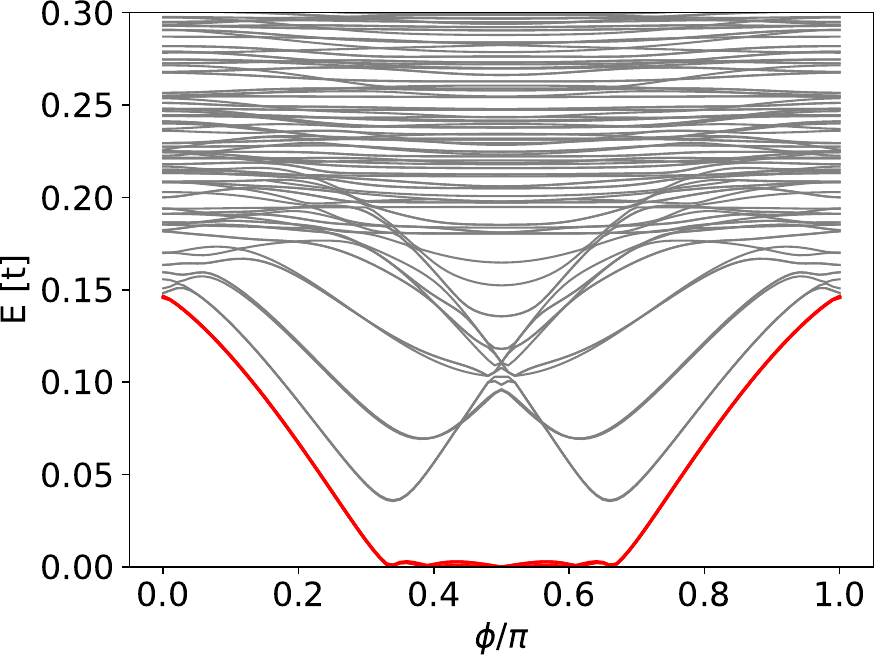}\tabularnewline
(d) & (e) & (f)\tabularnewline
\hspace{0.15cm} \includegraphics[width=4.5cm]{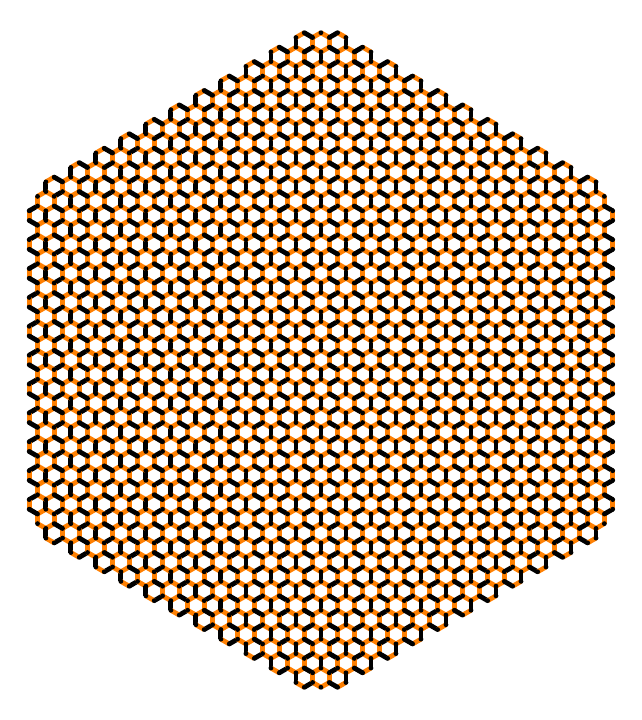} & \hspace{0.15cm} \includegraphics[width=4.5cm]{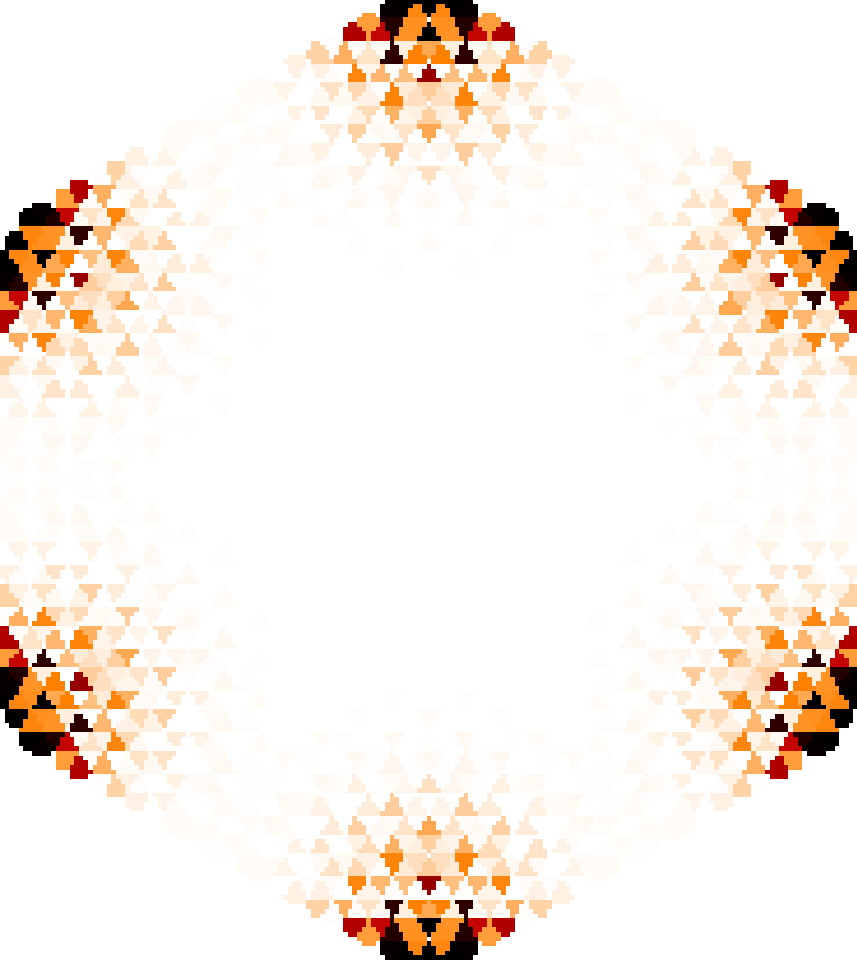} & \includegraphics[width=6cm]{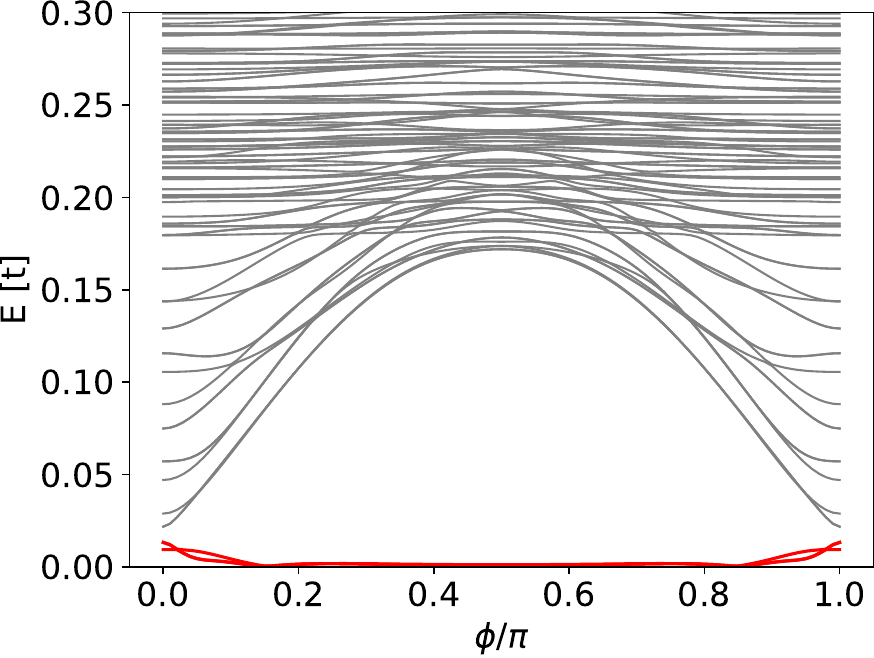}\tabularnewline
(g) & (h) & (i)\tabularnewline
\hspace{-0.2cm}\includegraphics[width=5.5cm]{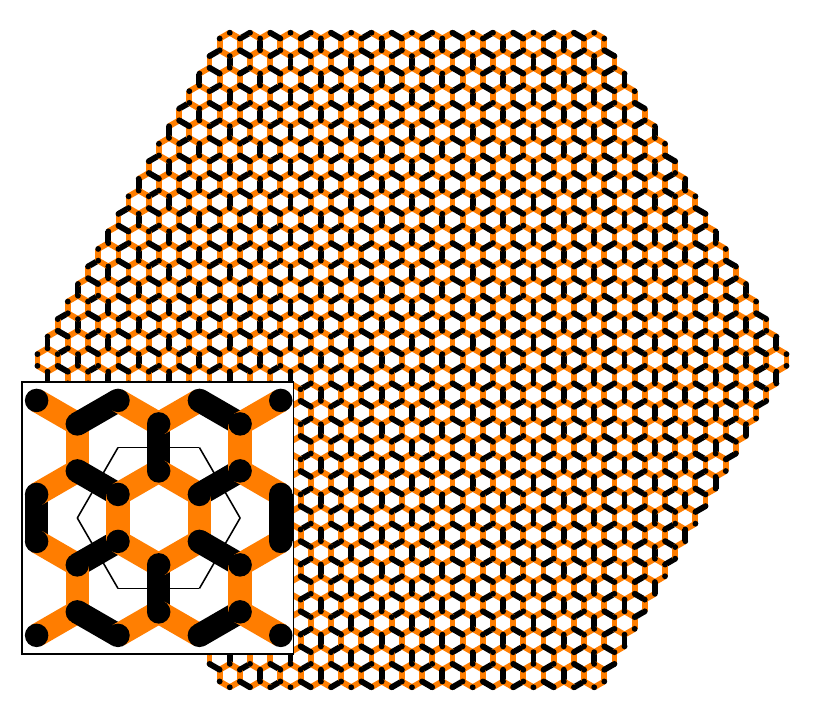} & \hspace{-0.2cm}\includegraphics[width=5.5cm]{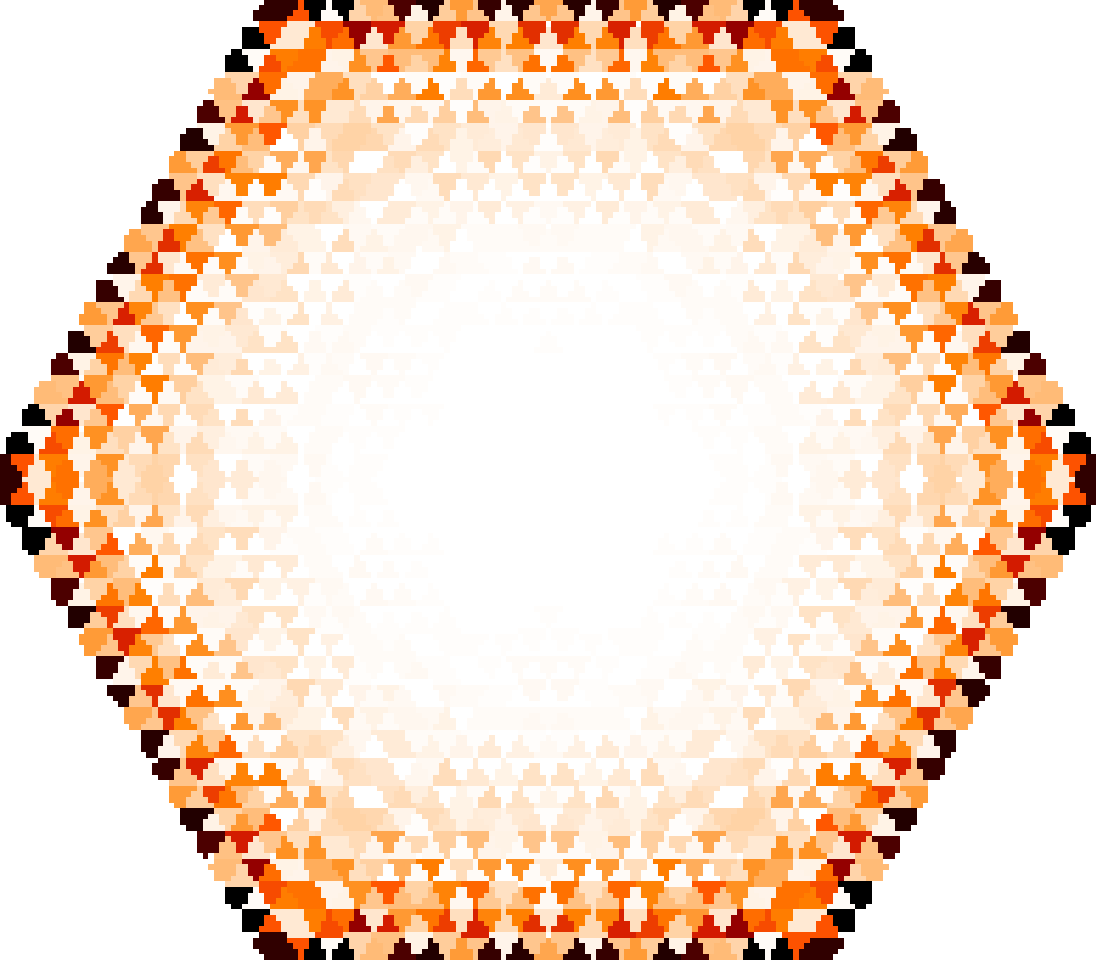} & \includegraphics[width=6cm]{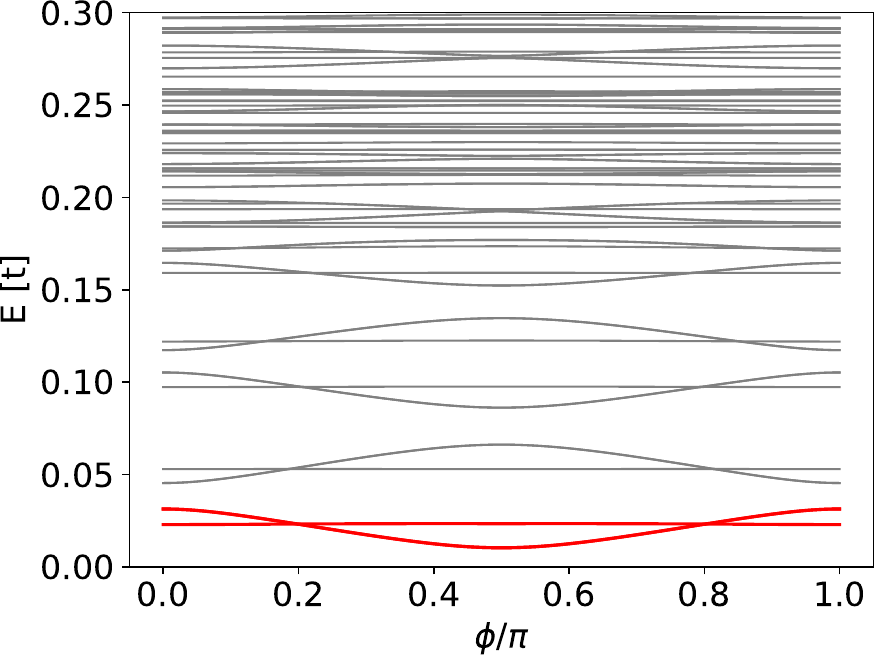}\tabularnewline
\includegraphics[width=5cm]{colorbar_lat} &  & \tabularnewline
\end{tabular}\caption{Results of the exact diagonalization of the Bogoliubov-de Gennes Hamiltonian within a single spin sector for hexagonal flakes of a honeycomb
lattice of three different geometries. The leftmost figures show the
spatially modulated pairing between nearest neighbors for $\phi=\pi/2$.
The color of the bonds between neighboring sites $\bm{r},\bm{r}'$
shows the value of the gap function $\Delta(\bm{r},\bm{r}')$ according
to the color scale. The inset of the bottom left figure depicts a
zoom into the bulk of the lattice with the unit cell boundary indicated
by the thin lines. The central figures show the wavefunction densities
of the three lowest quasiparticle modes. The rightmost figures show
the quasiparticle spectrum as a function of the pair density wave
parameter $\phi$. The three lowest quasiparticle modes are indicated
in red. For the exact diagonalization we use the parameters $\mu = 0.4t,\ \Delta^\prime = 0.2t$.}
\label{fig:ED:honeycomb}
\end{figure*}

Our RPA results show that the dominant superconducting instability for both the honeycomb and Kagome lattice is the intravalley $p$--wave spin triplet phase, with gap structure $\Delta_{\bm{k}} = |\Delta| e^{i\tau_z(\phi-\theta_{\bm{k}})} (d^\mu s_\mu)(i\tau_y)$. This gap possesses a number of unconventional features that distinguish it from the superconducting phases more commonly studied in either high-$T_c$ superconductors or Dirac materials. The gap exists at both $K$ points and has a $p+i\tau p$ wave angular structure, which points to the presence of a $\pi$--Berry phase in two-particle scattering from which superconducting pairing originates. However, the winding of the gap is opposite in the two valleys, with the superconducting order parameter at the two $K$ points being related to each other by complex conjugation. As a result, the usual topological features associated with chiral $p$-wave superconductors are strongly modified. We may view the phase as consisting of four copies of a topological $p+ip$ superconductor \cite{ReadGreen2000}, one per spin and valley species, with the gaps at opposite valleys possessing topological invariants of opposite sign. This results in a topologically trivial superconductor according to the Altland-Zirnbauer classification. However, in the presence of crystalline symmetries, a second order topology emerges which is characterized by anomalous corner modes.

In the intravalley paired state, Cooper pairs carry a finite quasimomentum, and as a result the gap is spatially modulated, forming a pair density wave which is commensurate with the lattice with a periodicity of three unit cells \cite{Fulde1964,Larkin1965,Roy2010,Tsuchiya2016,Honerkamp2008,Ma2011,Wu2013,Vladimirov2019}. The condensate possesses a U(1) order parameter $\phi$ associated with the spatial position of the pair density wave and does not couple to the electromagnetic field; the system also exhibits Goldstone modes associated with fluctuations of $\phi$, which are physically the sliding modes of the pair density wave. The spatial modulation of the gap spontaneously breaks the crystalline symmetries of the normal state except at special values $\phi = \frac{n\pi}{2}$, $n\in \mathbb{Z}$; at these values the gap may be classified according to a second order topological invariant which predicts the number of Majorana corner modes of a finite sample of the superconductor which preserves the crystalline symmetries.

We study the topology of the $p+i\tau p$ intravalley spin triplet phase via the bulk boundary correspondence for higher order topological superconductors, and identify the anomalous boundary physics in hexagonal superconducting flakes. By choosing a spin quantization axis which is aligned so that $\bm{d}\parallel \hat{\bm{y}}$, we find that pairing occurs between electrons with both spins aligned either along the $z$ or $-z$ axis. Thus the pairing term may be decomposed into a sum of two spin sectors. We derive the topological features of the superconducting state by analyzing the spinless Hamiltonian corresponding to pairing within a single spin sector. In order to make numerical diagonalization easier we employ a simplified form of the gap function in which we only include nearest neighbor pairing and hopping terms.  In the Appendix, we derive the real space form of the gap from the $p+i\tau p$ momentum space structure. The spinless mean field lattice Hamiltonian which results is 
\begin{gather}
H_{\text{BdG}} = \sum_{\langle \bm{r},\bm{r}'\rangle}{-tc^\dag_{\bm{r}} c_{\bm{r}'} + \frac{1}{2}\left[
\Delta(\bm{r},\bm{r}') c^\dag_{\bm{r}}c^\dag_{\bm{r}'} +\text{H.c.}\right]} 
\ \ ,
\label{Hamiltonian:BdG}
\end{gather}
where $\Delta(\bm{r}',\bm{r}) = -\Delta(\bm{r},\bm{r}')$, and 
\begin{gather}
\label{realDK}
\Delta(\bm{r},\bm{r}') = 
\Delta'\cos(\bm{K}\cdot(\bm{r}+\bm{r}') + \phi) \ , \ \begin{cases}\bm{r} \in B, \ \bm{r}'\in C \\
\bm{r}\in B, \ \bm{r}'\in A \\
\bm{r}\in A,  \ \bm{r}'\in C
\end{cases}
\end{gather}
for the kagome lattice, and
\begin{gather}
\label{realDH}
\Delta(\bm{r},\bm{r}') = \Delta'\sin(\bm{K}\cdot(\bm{r}+\bm{r}') + \phi) \ \ ,  \ \bm{r}\in B, \bm{r}'\in A
\end{gather}
for the honeycomb lattice, and we distinguish the parameter $\Delta'$ from the bulk gap $\Delta$.

We present exact diagonalization results for the kagome and honeycomb lattices in Figs. \ref{fig:ED:kagome}, \ref{fig:ED:honeycomb} respectively. We have calculated the spectrum for all values of $\phi$, however the topological classification must be performed at the values $\phi = 0, \frac{\pi}{2}$ for which the gap obeys a twofold rotation symmetry. We have performed calculations for a variety of edge geometries, and in each case we preserve the twofold rotational symmetry of the bulk. In a number of cases we observe a gapless edge spectrum for all values of $\phi$. This occurs for the geometries shown in the last two rows of Fig. \ref{fig:ED:kagome} and the last row of Fig. \ref{fig:ED:honeycomb}. These surfaces states are remnants of the chiral Majorana edge modes expected from the Read-Green model \cite{ReadGreen2000} which are the most easily identifiable manifestions of the chiral $p$-wave nature of the gap. For other edge terminations, however, the edge spectrum is gapped, indicating that the gapless edge modes are not topologically protected, which is consistent with the fact that, since our gap lies in class BDI, it cannot possess a first order topological invariant. For the other edge terminations, we find that at $\phi = \frac{\pi}{2}$, the edge states are gapped, and six zero energy modes  appear within each spin sector which are localized at each of the corners of the flake,  forming the three lowest modes within the quasiparticle spectrum of a single spin sector highlighted in red in Figs. ~\ref{fig:ED:kagome},~\ref{fig:ED:honeycomb}c,f,i. This occurs for both the honeycomb and kagome lattices. At $\phi = 0$, we either find no Majorana zero modes when the edge spectrum is gapped or a gapless edge spectrum. This allows us to conclude that both the honeycomb and kagome lattices are in a second order topological phase for $\phi = \frac{\pi}{2}$ and trivial for $\phi = 0$. We note that, since our results are for a spinless model obtained after performing a spin decomposition of the mean-field Hamiltonian, the Majorana zero modes appear as Kramers pairs. Since pairing occurs between electrons with parallel spin, the Majorana modes individually possess  definite spin, and are protected against hybridization by a spinful time reversal symmetry $\mathcal{T}' = e^{i\pi S_y} \mathcal{T}$.

Since the boundary modes in a higher order topological phase require crystalline symmetries, they are not protected against local perturbations. The robustness of the bulk-boundary correspondence in the presence of crystalline symmetry breaking therefore provides an important, general question for systems exhibiting higher order topology. In our examples, for general values of the pair density wave order parameter $\phi$, all the point group symmetries of the normal state are spontaneously broken  by the pairing term except for mirror reflection about the $x$ axis, which is preserved due to the fact that the phase $\bm{K}\cdot(\bm{r}+\bm{r}') + \phi$ of the pair density wave does not depend on the $y$ coordinate. However, at discrete values $\phi = \frac{\pi n}{2}$, the superconductor recovers a twofold rotational symmetry. The parameter $\phi$ allows continuous tuning of the system between a second order topological phase at $\phi = \pi(n+\frac{1}{2})$ and a trivial phase at $\phi = n\pi $. We observe, despite the breaking of crystalline symmetries, that the anomalous corner modes persist and remain exponentially close to zero energy in a finite region of values around $|\phi - \pi(n+\frac{1}{2})| < \phi^*$ up to a critical value $\phi^*$ at which the edge gap closes. The corner modes are protected by both the bulk and edge gap, and thus persist even when the point group symmetry is weakly broken \cite{Langbehn2017, Geier2018}.

The anomalous corner modes coexist with a gapped edge spectrum. This originates from the $p+i\tau p$ nature of the gap, which provides a 1D boundary superconductor for each copy of the Moore-Read model existing in each spin and valley species. The manifestation of the bulk second order topology is shadowed by the behavior of the boundary superconductor: for certain edge terminations, the edge modes remain anomalous due to symmetries along the edge which forbid their hybridization.  In particular, for the bottom geometries of Fig. \ref{fig:ED:kagome} and \ref{fig:ED:honeycomb}, the system possesses a mirror symmetry at $\phi = \frac{\pi n}{2}$ that prohibits the counterpropagating chiral Majorana modes from gapping out. In this case, the edge states are anomalous and associated with a bulk crystalline symmetry protected topological phase protected by mirror symmetry. For other edge terminations, mixing of the chiral modes between opposite valleys along the same edge occurs, and the edge spectrum is gapped. Interestingly, the emergence of anomalous corner modes in this situation is not general but depends on the value of $\phi$.

We  may make an interesting comparison to the $p+ip$ intervalley spin triplet phase, which is subleading in our analysis in the regime where the onsite couplings dominate over the nearest neighbor couplings, but may arise as the dominant instability for models with more complicated lattice interactions. In this case, pairing occurs between the electron pockets surrounding opposite valleys, and the order parameter spontaneously chooses a winding number corresponding to either $p+ip$ or $p-ip$ pairing. Time reversal symmetry is broken, which puts the superconductor in class D, rather than BDI as in the intravalley case. This phase possesses a Chern number which leads to a gapless, anomalous edge spectrum. In a forthcoming study, we show that this phase may become dominant when spin orbit coupling is present, and analyze the topology of this phase in detail.

\section{Discussion}
\label{disc}
Our results show that the first order topological $p+ip$ and second order topological $p+i\tau p$ superconducting states emerge naturally in the presence of Dirac points near the chemical potential. While in a previous study we explored the pseudospin pairing mechanism giving rise to a $p+i\tau p$ instability in a semiconductor-based artificial honeycomb lattice within the RPA  \cite{Li2020}, in this work we have extended our analysis of the pairing mechanism to generic lattices satisfying $C_{6v}$ symmetry and shown that this instability is not specific to any lattice structure but rather emerges due to the universal properties of the Dirac fermion excitations in these materials. We have illustrated that both the Berry phase and screening effects underlying the pairing mechanism are active in kagome lattices, and presented exact diagonalization results which demonstrate the striking similarity between the boundary physics in kagome and honeycomb lattices, which result from the fact that they possess identical topological properties.

In addition to the $p+i\tau p$ phase, the intervalley $p+ip$ and $s$-wave phase are possible, but are subleading in our analysis. Unlike the $p$-wave states, the $s$-wave state is protected against the effects of weak disorder by a generalized Anderson's theorem, so this state might posses a higher $T_c$ than the topological phases in a sufficiently disordered system. In a forthcoming study we also suggest that this phase may arise due to spin-orbit coupling and study its properties in detail.

We call attention to a number of relevant features of this mechanism: ($i$) the mechanism can be applied for weak repulsive interactions, so strong correlations are not necessary, ($ii$) superconductivity appears beyond a critical doping beyond the Dirac points, at which the screening effects become strong enough to give rise to a net attraction, ($iii$) the electronic origin of the pairing means the cutoff which determines $T_c$ is the Fermi energy rather than the Debye frequency, ($iv$) superconductivity is more pronounced in systems with localized orbitals.

Superconductivity has recently been discovered in quasi two dimensional vanadium-based kagome metals AV$_3$Sb$_3$ (A=K,Rb,Cs) \cite{Ortiz2020,Ortiz2021}. The agreement between DFT calculations of the bandstructure and ARPES suggests that these materials are weakly correlated, an interpretation also supported by the results of DFT/DMFT calculations \cite{Ortiz2020,Zhao2021,Tan2021}. The Fermi surface consists of two pockets surrounding the Dirac points, a pockets surrounding the $\Gamma$ point and an additional Fermi contour \cite{Kang2021}. Superconductivity exists alongside density wave order which sets in at a higher temperature and appears to compete with superconductivity \cite{Jiang2021,Li2021,Ortiz2019,Zhao2021b,Li2021b,Qian2021,Christensen2021}. STM and Josephson STS measurements suggest a spatially modulated superconducting gap \cite{Chenb2021} and zero bias peaks inside magnetic field-induced vortex cores are suggestive of Majorana bound states \cite{Liang2021}.

Given that phonons appear unable to account for the measured $T_c$ \cite{Chenb2021,Tan2021} and yet the materials appear to be weakly correlated, we argue that our mechanism might provide an explanation for superconductivity in kagome metals, in contrast to existing theoretical proposals which attribute superconductivity to the effects of nesting or competing density wave order \cite{Park2021,Wu2021,Lin2021}. In this interpretation, superconductivity originates from the Dirac-like Fermi surfaces and is not directly related to the observed charge density wave order \cite{Jiang2021}, which is potentially due to van Hove singularities or nested portions of the band structure near the Fermi level. Future theoretical work should investigate the interplay between the pairing mechanism and the presence of density wave order, which breaks rotational symmetry \cite{Li2021}. Future experiments could look for evidence of the edge or corner modes present in the $p+ip$ and $p+i\tau p$ phases \cite{Gray2019,Choi2020}, probe the spin structure of the gap through NMR, or further investigate the real space structure of the superconducting gap.

Additional kagome systems have recently been discovered to host superconductivity, and might also be possible candidates for the pseudospin mechanism, including bilayer kagome systems \cite{Baidya2020}, ferromagnetic kagome metals \cite{Ye2018}, and lanthanum-based materials \cite{Mielke2021}. Superconductivity has also been seen in Dirac surface states of doped topological insulators \cite{Bian2016, Neupane2016, Wray2010, Zhiwei2016, Han2015}, which might also be explained by this theory. In materials such as transition metal dichalcogenides like MoS$_2$ and MoTe$_2$, and few-layer stanene \cite{Liao2018}, the presence of spin orbit interactions gaps the Dirac points and/or introduces a valley-Zeeman field, but the Dirac physics persists nonetheless. In a forthcoming paper, we have examined the effects of spin-orbit coupling on the possible superconducting states, and find that the pseudospin mechanism results in topological superconductivity in these systems as well. As a general guiding principle, promising candidate materials for this mechanism are those with strongly localized orbitals -- which lead to larger values of the couplings $g_i$ and lower values of the critical doping $\mu^*$ -- and larger Dirac-like Fermi surfaces -- which enhance the screening effects.

\section{Acknowledgements}

The authors thank P. Brouwer, M. Scheurer and P. Kotetes for helpful discussions.

MG acknowledges support by the European Research Council (ERC) under the European Union’s Horizon 2020 research and innovation program under grant agreement No.~856526, and from the Deutsche Forschungsgemeinschaft (DFG) Project Grant 277101999 within the CRC network TR 183 "Entangled States of Matter" (subproject A03 and C01), and from the Danish National Research Foundation, the Danish Council for Independent Research | Natural Sciences.  H. D. Scammell acknowledges funding from ARC Centre of Excellence FLEET. TL acknowledges support from the DFG within the CRC network TR 183.

\let\oldaddcontentsline\addcontentsline
\renewcommand{\addcontentsline}[3]{}

\let\addcontentsline\oldaddcontentsline

\widetext
\newpage
\begin{center}
\textbf{\large Appendices}
\end{center}

\setcounter{equation}{0}
\setcounter{table}{0}
\setcounter{page}{1}
\setcounter{section}{0}
\makeatletter
\renewcommand{\theequation}{A\arabic{equation}}
\renewcommand{\thefigure}{A\arabic{figure}}
\renewcommand{\thesection}{A\arabic{section}}
\renewcommand{\bibnumfmt}[1]{[A#1]}
\renewcommand{\citenumfont}[1]{A#1}

\section{Symmetry relations for the effective field theory}

We may explicitly derive the interaction parameters $\mathcal{V}_{ab}$ in Eq. \ref{Hamiltonian_density} of the main text by projecting the extended Hubbard interactions onto the valley and pseudospin eigenstates $u_{\tau\alpha}(\bm{r})$. We find
\begin{gather}
\sum{\mathcal{V}_{ab} J^a_{\tau_1\tau_3;\alpha_1\alpha_3} J^b_{\tau_2\tau_4;\alpha_2\alpha_4}} = \sum_{\bm{r}_i}{U(\bm{r}_1,\bm{r}_2,\bm{r}_3,\bm{r}_4) u^*_{\tau_1\alpha_1}(\bm{r}_1) u_{\tau_3\alpha_3}(\bm{r}_3) u^*_{\tau_2\alpha_2}(\bm{r}_2) u_{\tau_4\alpha_4}(\bm{r}_4)} \ \ .
\label{ints:symmetry}
\end{gather}
The Hubbard interactions may be expressed in terms of the Wannier orbitals $\phi_{\bm{r}}(x)$ (with $x$ being a 3D coordinate vector) via
\begin{gather}
U(\bm{r}_1,\bm{r}_2,\bm{r}_3,\bm{r}_4) = \int{V(x-x') \phi_{\bm{r}_1}(x) \phi_{\bm{r}_3}(x) \phi_{\bm{r}_2}(x') \phi_{\bm{r}_4}(x') d^3 x d^3 x'} \ \ ,
\end{gather}
where $V(x-x')$ is the Coulomb interaction, and we have chosen the Wannier orbitals to be purely real. The interactions thus satisfy $U(\bm{r}_1,\bm{r}_2,\bm{r}_3,\bm{r}_4) = U(\bm{r}_3,\bm{r}_2,\bm{r}_1,\bm{r}_4)$.

We may use the representation of time reversal symmetry (Eq. (\ref{rep:symmetries}) in the main text) $\mathcal{T}= \Omega \mathcal{K}$ with $\Omega = \tau_x\alpha_x$ to derive a relation
\begin{gather}
u^*_{\tau\alpha}(\bm{r}) = \sum_{\tau',\alpha'}{ u_{\tau'\alpha'}(\bm{r}) \Omega_{\tau'\tau;\alpha'\alpha}} \ \ .
\end{gather}
Applying this to the factors $u_{\tau_1\alpha_1}(\bm{r}_1)$ and $u_{\tau_3\alpha_3}(\bm{r}_3)$ in (\ref{ints:symmetry}) we find
\begin{gather}
\sum{\mathcal{V}_{ab} J^a_{\tau_1\tau_3;\alpha_1\alpha_3} J^b_{\tau_2\tau_4;\alpha_2\alpha_4}} 
= \Omega_{\tau'_1\tau_1;\alpha'_1\alpha_1} \Omega^*_{\tau'_3\tau_3;\alpha'_3\alpha_3} \sum{\mathcal{V}_{ab} J^a_{\tau'_3\tau'_1;\alpha'_3\alpha'_1} J^b_{\tau_2\tau_4;\alpha_2\alpha_4}} \ \ .
\end{gather}
We may express this in the form
\begin{gather}
\sum_{ab}{}\mathcal{V}_{ab} J^a\otimes J^b = \sum_{ab}{}\mathcal{V}_{ab} (\Omega^\dag  J^a \Omega)^T \otimes J^b \ \ .
\end{gather}
Since the adjoint basis elements $J^a$ are Hermitian and transform with definite sign under time reversal, $(\Omega^\dag J^a \Omega)^T = (\Omega^\dag J^a \Omega)^* = \mathcal{T}^{-1} J^a \mathcal{T} = \pm J^a$, we find that $\mathcal{V}_{ab} = 0$ for any operator $J^a$ which is odd under time reversal.

We may derive further symmetry constaints using the relations
\begin{gather}
u_{\tau\alpha}(\Lambda^{-1}\bm{r}) = u_{\tau'\alpha'}(\bm{r}) U^\Lambda_{\tau'\tau;\alpha'\alpha}
\end{gather}
where $\Lambda$ is a symmetry operation acting on coordinates and $U^\Lambda$ is its unitary representation. Applying a simultaneous transformation $\bm{r}_i \rightarrow \Lambda \bm{r}_i$ in (\ref{ints:symmetry}) we find
\begin{gather}
\sum_{ab}{\mathcal{V}_{ab} J^a \otimes J^b } = \sum_{ab}{\mathcal{V}_{ab} ((U^\Lambda)^\dag J^a U^\Lambda)\otimes ((U^\Lambda)^\dag J^b U^\Lambda)} \ \ .
\end{gather}
Since there always exists a symmetry operation under which $J^a$ and $J^b$ transform with opposite signs unless $a=b$, we find that $\mathcal{V}_{ab}$ is a diagonal matrix.
\newpage

\section{Landau-Ginzburg analysis of the superconducting gap symmetry}
\label{LGanalysis}

The generic mean field Hamiltonian to account for all pairing possibilities, in the upper band, is

\begin{align}
\label{bdg_supp}
H_{BdG}&=\sum_{\bm k, s, \tau}\varepsilon_{\bm k} f^\dag_{\bm k s \tau}f_{\bm k s \tau} + \frac{1}{2}\sum_{\bm k,s, \tau,s', \tau'} f^\dag_{\bm k s \tau} \left(\Delta_{\bm k} i s_y i \tau_y\right)_{s \tau, s' \tau'} f^\dagger_{-\bm k s' \tau'} + \text{h.c.} -\frac{1}{2}\sum_{\bm k,\bm p}\Delta_{\bm k}^\dagger \Gamma^{-1}(\bm k ,\bm p)\Delta_{\bm p}
\end{align}
The gap takes any of the forms presented in Table \ref{tab:gaps}, which contains the valley, spatial and spin structure. The spin structure employs standard notation of $d^\mu$-vectors; 
\begin{align}
d^x&=\frac{1}{2}\left(\ket{\uparrow\uparrow}-\ket{\downarrow\downarrow}\right), \ \ d^y=\frac{1}{2i}\left(\ket{\uparrow\uparrow}+\ket{\downarrow\downarrow}\right), \ \
 d^z=\frac{1}{2}\left(\ket{\uparrow\downarrow}+\ket{\downarrow\uparrow}\right), \  \ d^0=\frac{1}{2}\left(\ket{\uparrow\downarrow}-\ket{\downarrow\uparrow}\right).
\end{align}

The analysis presented in the main text states that the gap functions can be chosen as simply one of the possible degenerate sets; for the gap structure discussed in Section IV it is important that we can choose the $d$-vector along one axis $\Delta\propto d_y s_y$, and that a non-unitary choice of $\Delta_{\bm k}$ is energetically penalized. We explicitly show that this is the case below. We compute the free energy to quartic order from \eqref{bdg_supp}, and find
\begin{align}
{\cal F}&= \frac{1}{2}\sum_{\bm k,\bm p}\Delta_{\bm k}^\dagger \Gamma^{-1}(\bm k ,\bm p)\Delta_{\bm p}+\frac{1}{4}\Tr \left[\frac{\Delta_{\bm k} \Delta_{\bm k}^\dag}{\omega^2+\epsilon_k^2}\right]  + \frac{1}{8}\Tr \left[\frac{\Delta_{\bm k} \Delta_{\bm k}^\dag\Delta_{\bm k} \Delta_{\bm k}^\dag}{(\omega^2+\epsilon_k^2)^2}\right] 
\end{align}
where $\Gamma$ are the Cooper channel Born amplitudes, explicitly given in \eqref{BCS:vertex}, and the trace includes summation over momentum and frequency.

Each gap structure entering Table \ref{tab:gaps} has a critical temperature set by the coupling $\lambda$, where $\lambda = \Gamma^\ell_{\tau\tau\tau\tau}$ for intravalley pairing, while $\lambda = \Gamma^{\ell}_{+-+-} \pm \Gamma^{\ell}_{+--+}$ for intervalley pairing. We restrict our attention to work within each degenerate set of gaps separately. In most cases, the valley structure is trivially traced out -- the free energy reduces to separate cases based on the overall spin degeneracy $\bm d\cdot \bm s$ and degenerate chiral phase factors $e^{\pm i\theta_{\bm k}}$. Using this logic, there are only five distinct cases to consider 
\begin{align}
\notag (i) \ \ &d^0 s_0 \ \tau_z \\
\notag (ii) \ \ &d^0_{\pm} s_0 \ e^{\pm i \theta_{\bm{k}}} \tau_0 \\
\notag (iii) \ \ & \bm{d}\cdot \bm{s} ,\ \ \bm{d}\cdot \bm{s}\ e^{i\tau_z(\phi-\theta_{\bm{k}})} \ (i\tau_y)\\
(iv) \ \ &  \bm{d}_\pm\cdot \bm{s} \ e^{\pm i \theta_{\bm{k}}} \ \tau_z \nonumber \\
(v) \ \ &  d^0 s_0 \left(1,  e^{\pm i \theta_{\bm{k}}} , e^{-2i\tau_z \theta_{\bm{k}}}e^{i\tau_z \theta_{\bm{k}}} \right) e^{i\tau_z\phi}(i\tau_y) \equiv (v_0, v_{-2\tau}, u_{\tau\pm1})
\end{align}
Case $(iii)$ enumerates the two physically distinct valley structures that share the same spin structure and chirality, but themselves are not degenerate, i.e. have different $T_c$. The differing valley structures of these gaps does not affect the contribution to $\mathcal{F}$ that is quartic in $\Delta$;
the valley structure enters through $\Gamma$ and affects the value of $T_c$ but not the stability analysis. The brackets in $(v)$ accounts for the four physically distinct gap structures which are degenerate, i.e. have the same value of $\lambda$ and hence have same critical temperature, c.f. Table \ref{tab:gaps}. We combine these three gap functions into a vector $(v_0,v_{-2\tau},u_{\tau\pm 1}) $, which will use below in the free energy analysis.  

Evaluating the free energy to quartic order in the gaps,
\begin{align}
\notag {\cal F}_1[d^0 s_0]&= a_1(T-T_{c_1})|d_0|^2 + b  |d_0|^4 ,\\
\notag {\cal F}_2[ d^0_{\pm} s_0 \ e^{\pm i \theta_{\bm{k}}}]&= a_2(T-T_{c_2})(|d_{0+}|^2+|d_{0-}|^2) + b  \left(|d_{0+}|^4 + 4 |d_{0+}|^2 |d_{0-}|^2 +  |d_{0-}|^4\right) ,\\
\notag {\cal F}_3[ \bm{d}\cdot \bm{s} ]&= a_3(T-T_{c_3})(\bm d^*\cdot \bm d) + b \left\{(\bm d^*\cdot \bm d)^2+|\bm d^*\times \bm d|^2\right\} ,\\
\notag {\cal F}_4[ \bm{d}_\pm\cdot \bm{s} \ e^{\pm i \theta_{\bm{k}}} ]&= a_4(T-T_{c_4})(\bm d_+^*\cdot \bm d_+ + \bm d_-^*\cdot \bm d_-) + b \Big\{(\bm d_+^*\cdot \bm d_+)^2+(\bm d_-^*\cdot \bm d_-)^2+4(\bm d_+^*\cdot \bm d_+)(\bm d_-^*\cdot \bm d_-)\\
\notag &+|\bm d_+^*\times \bm d_+|^2 + |\bm d_-^*\times \bm d_-|^2 + 4(\bm d_+^*\times \bm d_+)(\bm d_-\times \bm d_-^*)\Big\}\\
\notag {\cal F}_5[(v_0, v_{-2\tau}, u_{\tau\pm1})]&= a_5(T-T_{c_5})\left(|v_0|^2 + |v_{-2\tau}|^2+|u_{\tau+1}|^2+|u_{\tau-1}|^2\right) + b\Big\{|v_0|^4 + |v_{-2\tau}|^4+|u_{\tau+1}|^4+|u_{\tau-1}|^4 \\
\notag &+ 4|v_0|^2\Big( |v_{-2\tau}|^2+ |u_{\tau+1}|^2+ |u_{\tau-1}|^2\Big) +4|v_{-2\tau}|^2\Big(|u_{\tau+1}|^2+ |u_{\tau-1}|^2\Big) + 4|u_{\tau+1}|^2|u_{\tau-1}|^2
\Big\}
\end{align}
with coefficients $a_i>0, b>0, i=1,2,3$. We summarize the main conclusions:
\begin{itemize}
    \item Case ($i$) is a $s$-wave singlet paired state. The free energy straightforwardly establishes stability of this phase. 
    
    \item  Case ($ii$) corresponds to an intervalley $p+ip$ state, in which electrons from opposite valleys pair with a definite chirality. From ${\cal F}_2[ d^0_{\pm} s_0 e^{\pm i \theta_{\bm{k}}}]$, we find that coexistence of opposite chiralities is energetically penalized so that the system spontaneously chooses one chirality, e.g. $d^0_+ s_0 \ e^{+ i \theta_{\bm{k}}}$, therefore breaking time reversal. This state exhibit first order topology with gapless edge states.

    \item Case ($iii$) concerns two triplet paired states for which the free energy has the same form; from ${\cal F}_3[ \bm{d}\cdot \bm{s} ]$ we find that non-unitary paring is penalised, i.e. $\bm{d}$ is purely real (or purely imaginary); the $\bm{d}$ vector spontaneously chooses a direction, breaking spin $SU(2)$ symmetry, but the condensate has vanishing magnetization. The two superconducting states included in this case are: intravalley $p+i\tau p$, in which electrons undergo $p+ip$ pairing in one valley and $p-ip$ pairing in the other, and intervalley $s$-wave, in which electrons from opposite valleys undergo triplet $s$-wave pairing. Both states respect time-reversal symmetry.

    \item Case ($iv$) examines spin-triplet $p$-wave intervalley pairing.  From ${\cal F}_4[ \bm{d}_\pm\cdot \bm{s} \ e^{\pm i \theta_{\bm{k}}} ]$, similar to case ($ii$), the resulting state spontaneously breaks time reversal symmetry, resulting in a first order topological $p+ip$ state e.g. $\bm d_+e^{+ i \theta_{\bm{k}}}$. As in case ($iii$), we again find the $d$-vector is purely real (or imaginary), and hence this state breaks time reversal symmetry but does not support a magnetization. 
    
    \item Case ($v$) looks at a degenerate manifold of $s$-wave, $d+i\tau d$, and a mixed state with $s$-wave in one valley and a $d+i d$ pairing in the opposite valley, the latter of which breaks time reversal. The intravalley $d$-wave state is a $d+i\tau d$ superconductor -- a time reversal invariant combination of $d+id$ pairing in one valley, and $d-id$ pairing in the other -- and exhibits second order topology with gapless corner states.

\end{itemize}

\newpage
\section{Screened interactions}
\label{screenedint}

In this appendix we discuss the screened electron-electron interactions. We begin by presenting general expressions for the polarization operators, and making some general comments about regulating UV divergences in effective quantum field theories. We then explicitly calculate the intra and inter valley current-current susceptibilities. We present the formulae for the screened couplings, and plot the frequency dependence of the resulting scattering amplitudes for Cooper pairs. We conclude by discussing alternative regularizations for computing the susceptibilities.

\subsection{Preliminaries}
The RPA equations for the screened interactions are
\begin{align}
\widetilde{V}_{\mu \nu}(\omega, \bm{q}) = 
 V_{\mu \nu}(\bm{q}) + V_{\mu \alpha}(\bm{q}) \Pi^{\alpha \gamma} (\omega,\bm{q}) \widetilde{V}_{\gamma \nu} (\omega, \bm{q})
\end{align} where $\Pi^{\alpha \gamma}$ is the polarization operator, 
\begin{gather}
\Pi^{\mu\nu}(\omega, \bm{q}) = 
-i \text{Tr} \int{
J^\mu G(E+\omega, \bm{k}+\bm{q}) J^\nu G(E, \bm{k}) \frac{ dE d^2\bm{k}}{(2\pi)^3}
} \ \ , \nonumber \\
G(E, \bm{k}) = \frac{1}{ E +\mu - v \tau_z \bm k \cdot \bm \alpha +i 0 \text{sgn}(E)}
\label{Pi}
\end{gather} 
where the vertices $J^\mu = \{J_1,\dots, J_{10}\}  = \{\tau_0 \alpha_0, \tau_z\alpha_z, \tau_0\alpha_x, \tau_0\alpha_y, \tau_x\alpha_0, \tau_y\alpha_0, \tau_x \alpha_x, \tau_x\alpha_y, \tau_y \alpha_x,\tau_y \alpha_y\}$, as per Section \ref{RPA} of the main text. We choose units in which $v=1$, and upon performing the frequency integral by residues, manipulations presented in \cite{Li2020b} result in
\begin{gather} 
\Pi^{\mu\nu}(\omega,q)= \text{Tr}\int \frac{d^2k}{(2\pi)^2} \sum_{s=\pm} \frac{J^\mu (s\tilde{\omega}+k - \tau_z(\bm k - s\bm q )\cdot \bm \alpha)J^\nu(-k+ \tau_z\bm k \cdot \bm \alpha)}{2k((s\tilde{\omega}+k)^2-(\bm k - s\bm q)^2)}  \nonumber \\
+\frac{J^\mu (s\tilde{\omega}+k +\tau_z(-\bm k + s\bm q )\cdot \bm \alpha)J^\nu(k-\tau_z\bm k \cdot \bm \alpha)}{2k((s\tilde{\omega}+k)^2-(\bm k - s\bm q)^2)}  \Theta(\mu-k)
\end{gather} 
The first term -- the interband polarization operator, denoted $\Pi_+$ -- contributes when $\mu=0$, while the second term -- the intraband polarization operator, denoted $\Pi_-$ -- only contributes when $\mu\neq 0$. Explicitly, we write
\begin{align} 
\label{intraPi}
\Pi^{\mu\nu}_+(\omega,q)= \text{Tr}\int \frac{d^2k}{(2\pi)^2} \sum_{s=\pm}\frac{J^\mu (s\tilde{\omega}+k -\tau_z(\bm k - s\bm q )\cdot \bm \alpha)J^\nu(k-\tau_z\bm k \cdot \bm \alpha)}{2k((s\tilde{\omega}+k)^2-(\bm k - s\bm q)^2)}  \Theta(\mu-k)
\end{align} 
and
\begin{align}
\label{interPi}
\Pi^{\mu\nu}_-(\omega,q)= -i\text{Tr} \int \frac{d^2k}{(2\pi)^2}\frac{dE}{2\pi}\frac{J^{\mu}((\omega+E)e^{i0}+\tau_z(\bm k+\bm q)\cdot \bm \alpha)J^{\nu}(E e^{i0}+\tau_z\bm k\cdot \bm \alpha)}{(((\omega+E)e^{i0})^2-(\bm k+\bm q)^2)((E e^{i0})^2-k^2)}
\end{align}

The polarization operator $\Pi^{11}$ is the standard charge polarization operator for graphene 
\cite{Son2007,Wunsch2006,Hwang2007b}, which describes the screening of the density-density interaction $\alpha_0\tau_0\otimes \alpha_0\tau_0$.  The function $\Pi^{22}$ is the pseudospin polarization operator which describes the screening behaviour of the pseudospin-pseudospin interaction $\alpha_z\tau_z\otimes \alpha_z\tau_z$ which was first calculated in \cite{Li2020b} (denoted there as $\Pi^{zz;00}$ due to different notation and a difference choice of basis for the pseudospin states). The remaining functions describe the screening of the intra and intervalley chiral currents. The current-current susceptibility for graphene has been discussed in \cite{Principi2009,Scholz2011,Khalilov2015}, though to the best of our knowledge the intervalley current-current susceptibility has not been previously investigated. Below, it will be convenient for us to derive expressions for both functions, and present them in a form somewhat more general than those already in the literature.

All polarization operators may all be written in terms of four basic functions, for which we will introduce simple notation in what follows. Use the spacetime indices $J^\mu = (\tau_0\alpha_0,\tau_0\alpha_i)$ where  $i=x,y$, and denote $\Pi^{22}$ as $\Pi^{zz}$. Considering $J^\mu = \alpha_i \tau_0$ and $J^\nu = \alpha_j \tau_0$, ie the chiral current screening $\Pi^{33},\Pi^{34},\Pi^{44}$, we then write this function as $\Pi^{ij}$ and decompose into longitudinal and transverse parts:
\begin{align}
\Pi^{ij}(\omega, \bm q) = \Pi_\perp(\omega, \bm q)(q^2\delta^{ij} - q^i q^j) +\Pi_\parallel(\omega, \bm q) q^i q^j
\end{align}

The chiral current polarization operator obeys an important identity due to electromagnetic gauge invariance. In the Dirac theory, the electromagnetic current operator $\bm J = \bm \alpha$ so that the vector potential of electromagnetism $\bm A$ appears in the Hamiltonian as the perturbation $e\bm J \cdot \bm A$. Gauge invariance, or equivalently the conservation of charge, can be shown to imply the Ward identity,
\begin{align}
q_\nu\Pi^{\mu\nu}(\bm q, \omega) = 0
\end{align}
From this we find 
\begin{align}
\omega \Pi^{0i}(\omega, \bm q) -  q_j\Pi^{ji}(\omega, \bm q) = 0 \nonumber \\
\omega \Pi^{00}(\omega, \bm q) -  q_i\Pi^{0i}(\omega, \bm q) = 0 \nonumber
\end{align}
Combining these,
\begin{align}
\omega^2 \Pi^{00}(\omega, \bm q) =  q_i  q_j \Pi^{ij}(\omega, \bm q) = q^4 \Pi_\perp(\omega, \bm q) 
\end{align}
which gives us an expression for the longitudinal part of the current-current susceptibility in terms of the density-density response
\begin{align}
\Pi_\parallel(\omega, \bm q) = \frac{\omega^2}{q^4} \Pi^{00}(\omega, \bm q)
\label{ward}
\end{align}
Comparing our below results with the expression for $\Pi^{00}$ cite \cite{Li2020b} shows our expressions satisfy this relation. We shall also derive an expression for the transverse part of the susceptibility in the dimensional regularization scheme,
\begin{align}
\label{relation}
q^2\Pi_{\perp}(\omega, \bm q) =  \tfrac{q^2-\omega^2}{q^2}\Pi^{00}(\omega, \bm q) + \Pi^{zz}(\omega, \bm q)
\end{align}
which therefore allows the current-current screening operator to be written entirely in terms of the density and pseudospin responses.

As is common quantum field theories, the polarization operators are formally divergent quantities and require regularization. However, the physical origins of the divergences and the effects of regularization differ between the polarization operators. The dimensional regularization scheme has the effect of simply setting all UV contributions to zero, leaving just the effects of those degrees of freedom in the Dirac effective theory near the $K$-points. Since actual materials are UV-completed by a lattice, placing the theory on a lattice is a more physical regularization scheme, and gives the physical relation between the numerical values of the couplings computed by band structure methods and the couplings in the effective theory.

Firstly, the charge susceptibility $\Pi^{00}(\omega,\bm q)$ obeys the exact compressibility sum rule, $\Pi^{00}(\omega=0,\bm q\rightarrow 0)=-\nu_0$ where $\nu_0$ is the density of states at the Fermi level, which means that cutoff-dependent quantities never appear in any sensible regularization scheme. For small frequencies and momenta, the regularized polarization operator must be the same as that calculated in dimensional regularization. 

Second, the current susceptibilities $\Pi^{ij}(\omega,\bm q)$ obey the Ward identity discussed above, as a result of the fact that $\bm \alpha$ is the current operator in the Dirac theory, which also excludes cutoff-dependent contributions. However, away from the $K$-points, the current operator is no longer given by $\bm \alpha$, which will result in UV-dependent contributions to $\Pi^{ij}(\omega,\bm q)$. In the lattice regularization scheme, this means that one can make changes to the lattice -- for instance by modifying the dispersion near the $\Gamma$ point -- that will result in a different value for $\Pi^{ij}(\omega,\bm q)$, but these UV-contributions originate from physics away from the Dirac point where the Ward identity applies. Similar to $\Pi^{00}(\omega,\bm q)$, for small frequencies and momenta, the regularized $\Pi^{ij}(\omega,\bm q)$ must be the same in lattice and dimensional regularization.

The pseudospin susceptibility $\Pi^{zz}(\omega,\bm q)$, by contrast, obeys no such constraint and receives a constant and negative cutoff dependent contribution in a lattice regularization. We may distinguish between two types of cutoff dependence -- contributions which originate from physics near the Dirac point, and contributions which originate from physics away from the Dirac point. While only the latter affects $\Pi^{ij}(\omega,\bm q)$, as we explained above, both types of UV-contributions appear in $\Pi^{zz}(\omega,\bm q)$; one way to see this is to use a hard cutoff $\Lambda$ in the Dirac theory, where (restoring the Dirac velocity $v$) one finds $\Pi^{00}(\omega=0,\bm q\rightarrow 0)=-\Lambda/(2\pi v)+\nu_0$. Thus by solely modifying physics near the Dirac point -- for instance by changing the velocity -- these UV contributions which originate from physics near the Dirac point are changed. As a result, these contributions which appear in a lattice regularization are important even at small frequencies and momenta, and are described in the main text.

When working with an effective field theory, one always performs calculations in terms of UV-independent quantities. However, if one wishes to perform a microscopic calculation of the couplings $g_i$ in \eqref{Hamiltonian} -- by computing the wavefunctions of a material for e.g. through band structure diagonalization and then taking matrix elements of the Coulomb interaction -- the couplings which appear in the effective quantum field theory will be related to those values through lattice regularization, in which $\Pi^{ij}(\omega,\bm q)$ and $\Pi^{zz}(\omega,\bm q)$ receive contributions from physics away from the Dirac point, and $\Pi^{zz}(\omega,\bm q)$ receives an additional constant contribution from physics near the Dirac point.

\subsection{Intravalley current-current polarization operator}

We now turn to a calculation of $\Pi^{ij}(\omega,\bm q)$. We  begin by calculating the interband part through dimensional regularisation, using the formulae
\begin{gather}
\label{dimreg1}
\int \frac{d^dk}{(2\pi)^d} \frac{(k^2)^a}{(k^2+\Delta)^b} = \frac{1}{(4\pi)^{d/2}}\ \frac{\Gamma(\frac{d}{2}+a)\Gamma(b-a-\frac{d}{2})}{\Gamma(\frac{d}{2})\Gamma(b)}\ \Delta^{d/2+a-b} \\
\label{dimreg2}
\int \frac{d^dk}{(2\pi)^d} \frac{(k^2)^a}{(k^2-\Delta)^b} = i\ \frac{(-1)^{a-b}}{(4\pi)^{d/2}}\ \frac{\Gamma(\frac{d}{2}+a)\Gamma(b-a-\frac{d}{2})}{\Gamma(\frac{d}{2})\Gamma(b)}\ \Delta^{d/2+a-b}
\end{gather}
Focus first on the denominator in Eq. \eqref{interPi}. Using the Schwinger--Feynman parametrization,
\begin{align}
\frac{1}{AB} = \int_0^1\frac{dx}{(xA+(1-x)B)^2}
\end{align}
we write
\begin{align}
\frac{1}{((\omega+E)^2-(\bm{k}+\bm{q})^2)(E^2-k^2)}=\int_0^1 dx \ \frac{1}{((k+xq)^2-x(x-1)q^2)^2}
\end{align}
where we now use relativistic notation $l^\mu=(E,\bm{k})$, $p^\mu=(\omega,\bm{q})$, and $l^2=E^2-\bm{k}^2$, $p^2=\omega^2-\bm{q}^2$. Shifting $l\rightarrow l-xp$ (bear in mind that this will affect the numerator as well), and Wick rotating $E \rightarrow iE$, the expression becomes
\begin{align}
\int_0^1 dx \ \frac{1}{(l^2+x(x-1)p^2)^2}
\end{align}
The corresponding numerator (before Wick rotation) is 
\begin{align}
\text{Tr} \  J^\mu \left((1-x)\omega+E + [(1-x)\bm q+ \bm k]\cdot \bm \alpha \right) J^\nu \left( -x\omega + E +[-x\bm q+\bm k]\cdot\bm \alpha\right)
\end{align}
Now we substitute $J^\mu,J^\nu=\alpha_i,\alpha_j$ with $i,j=x,y$ and perform the pseudospin trace using the identity
\begin{align}
\label{traceid}
\text{Tr} \left[ \alpha_i (A + B^\lambda\alpha_\lambda) \alpha_j (C + D^\rho \alpha_\rho) \right] = 2 AC\delta^{ij} +2\left( B^i D^j + B^j D^i - B\cdot D \delta^{ij} \right)
\end{align}
where $\Delta=x(x-1)p^2$. Using rotational symmetry to replace $(l^i)^2\rightarrow\frac{1}{3}l^2$, we arrive at
 \begin{align}
 \label{dimregint}
 \Pi^{ij}_- = 2N\int_0^1 dx  \int \frac{d^dl}{(2\pi)^d} \frac{[x(x-1)p^2-\frac{1}{3}l^2]\delta^{ij}-2x(1-x) q^i q^j}{(k^2+\Delta)^2}
 \end{align}
 where we account for the valley and spin trace through a factor $N=4$. Integrating using Eq. \eqref{dimreg2}, we find the imaginary part 
 \begin{align}
 \text{Im} \ \Pi^{ij}_- = 2N\int_0^1 \frac{dx}{8\pi} \  \left\{ \left\{ \frac{x(x-1)p^2}{\sqrt{x(1-x)p^2}}-\sqrt{x(1-x)p^2} \right\}\delta^{ij} +\frac{2x(x-1)p^ip^j}{\sqrt{x(1-x)p^2}}\right\}\Theta(p^2)
 \end{align}
Using $\int_0^1 dx \sqrt{x(1-x)} = \pi/8$, we arrive at 
 \begin{align}
 \text{Im} \ \Pi^{ij}_- = -\frac{N}{16\sqrt{\omega^2-q^2}}\left\{\delta^{ij}(\omega^2-q^2) + q^i q^j \right\} \ \Theta(\omega-q)
 \end{align}
 Similarly using Eq. \eqref{dimreg1}, the real part is calculated from Eq. \eqref{dimregint} to be
 \begin{align}
 \text{Re} \ \Pi^{ij}_- = -\frac{N}{16\sqrt{q^2-\omega^2}}\left\{\delta^{ij}(\omega^2-q^2) + q^i q^j \right\} \ \Theta(q-\omega)
 \end{align}
The functions $\Pi^{ij}_\perp(\Omega, \bm q)$ and $\Pi^{ij}_\parallel(\Omega, \bm q)$ are respectively the transverse and longitudinal components of the polarization operator. From above, we see the their interband parts are 
\begin{align}
\label{intertrans}
 \Pi^{ij}_{-,\perp} = -\frac{N}{16 q^2} \sqrt{q^2-\omega^2} \ \Theta(q-\omega)+i\frac{N}{16 q^2} \sqrt{\omega^2-q^2} \ \Theta(\omega-q)
\end{align}
and
\begin{align}
 \ \Pi^{ij}_{-,\parallel} = -\frac{N}{16q^2}\frac{\omega^2}{\sqrt{q^2-\omega^2}} \ \Theta(q-\omega)-i\frac{N}{16q^2}\frac{\omega^2}{\sqrt{\omega^2-q^2}} \ \Theta(\omega-q)
\end{align}
Calculating the density response $J^\mu=J^\nu=\alpha_0$ through the formalism above, one may see that the above expression for $\Pi^{ij}_\parallel$ satisfies the Ward identity \eqref{ward}. 

Let us now perform the intraband calculation. The numerator in Eq. \eqref{intraPi} is evaluated using Eq. \eqref{traceid}, so that
\begin{align} 
\Pi^{ij}_+(\omega,q)= 2N\int \frac{d^2k}{(2\pi)^2} \sum_{s=\pm}\frac{ \{ k(s\widetilde{\omega}+k) - \bm k \cdot (\bm k- s \bm q)\}\delta^{ij} + 2k^ik^j - s(k^iq^j + k^jq^i)  }{2k((s\tilde{\omega}+k)^2-(\bm k - s\bm q)^2)}  \Theta(\mu-k)
\end{align} 
The denominator equals $\widetilde{\omega}^2 + 2s\widetilde{\omega}k -q^2+2 s \bm k \cdot \bm q$. Shifting the angle of integration $\theta_{\bm k } \rightarrow \theta_{\bm k } + \theta_{\bm q } $ where $\theta_{\bm q } $ is the angle of the vector $\bm q$, and denoting the angle of integration hereafter as simply $\theta$, we have $\bm k \cdot \bm q \rightarrow kq \cos \theta$. Using the evenness of the denominator in $\theta$, the numerator may be simplified to arrive at 
\begin{align} 
\label{simplif}
\Pi^{ij}_+(\omega,q)= 2N\int \frac{dk d\theta}{(2\pi)^2} \sum_{s=\pm}\frac{ \{ s\widetilde{\omega}k  + skq \cos \theta\}\delta^{ij} + 2k^2\cos\theta \delta^{ij} + 2k^2\cos 2\theta \hat{q}^i\hat{q}^j - 2skq\cos\theta\hat{q}^i\hat{q}^j   }{\widetilde{\omega}^2 + 2s\widetilde{\omega}k -q^2+2 s kq \cos\theta}  \Theta(\mu-k)
\end{align} 
where $\hat{q}^i = q^i/q$. To proceed, we make use of the integral
\begin{align}
\label{cos2}
 I(a + ib0)=\int_0^{2\pi} \frac{1}{a+ib0+\cos\theta} \frac{d\theta}{2\pi }= \frac{\text{sgn}(a)}{\sqrt{a^2-1}} \Theta(|a|-1)-i \frac{\text{sgn}(b)}{\sqrt{1-a^2}}\ \Theta(1-|a|)
\end{align}
and the identities
\begin{align}
\label{cos2}
\int_0^{2\pi} \frac{\cos\theta}{a+\cos\theta} \frac{d\theta}{2\pi}  &= 1- aI(a ) \\
\int_0^{2\pi} \frac{\cos 2\theta}{a+\cos\theta} \frac{d\theta}{2\pi}  &=  -2a + (2a^2-1)I(a )
\end{align}
Evaluating \eqref{simplif}, we define $\alpha = \frac{\widetilde{\omega}^2 + 2s\widetilde{\omega}k -q^2}{2kq} = \frac{{\omega}^2 + 2s{\omega}k -q^2}{2kq} + i 0\frac{{\omega}^2 + s{\omega}k}{kq}$ and denote $\tilde{I}(\alpha) = 2kqI(\alpha)$. Then
\begin{gather} 
\Pi^{ij}_+(\omega,q)= 2N\int \frac{dk }{2\pi} \sum_{s=\pm}\left( (\tfrac{1}{2}+ \tfrac{k\alpha}{q}+\{s\widetilde\omega k-kq \alpha+2k^2(1-\alpha_2)\}\tilde I(\alpha))\delta^{ij} \right. \nonumber \\
\left. + \ (-1+ \tfrac{2k\alpha}{q}+\{2k^2+2kq\alpha-4k^2(1-\alpha_2)\}\tilde I(\alpha) )\hat{q}^i\hat{q}^j  \right) \Theta(\mu-k)
\end{gather} 
Making the substitution $2p=2k+s\widetilde\omega$, we get
\begin{gather} 
\Pi^{ij}_+(\omega,q)= \int \frac{dk }{2\pi} \sum_{s=\pm}\left( (\tfrac{\widetilde{\omega}^2}{2q^2}+ \tfrac{1}{q^2}(q^2-\widetilde{\omega}^2)\tilde I(\alpha))\left[\delta^{ij}-\hat{q}^i\hat{q}^j\right]  - (\tfrac{\widetilde{\omega}^2}{2q^2}-\tfrac{\widetilde{\omega}^2}{2q^2}(4p^2-q^2) \tilde I(\alpha) )\hat{q}^i\hat{q}^j  \right) \Theta(\mu-k)
\end{gather} 
We arrive at
\begin{align}
\Pi^{ij}_+(\omega, \bm q) = \Pi_{+,\perp}(\omega, \bm q)(q^2\delta^{ij} - q^i q^j) +\Pi_{+,\parallel}(\omega, \bm q) q^i q^j
\end{align}
with 
\begin{align}
\Pi_{+,\perp} = 2N\int \frac{dk }{2\pi} \sum_{s=\pm}\left( \tfrac{\widetilde{\omega}^2}{2q^4}+ \tfrac{1}{q^4}(q^2-\widetilde{\omega}^2)\tilde I(\alpha) \right) \Theta(\mu-k)
\end{align}
and
\begin{align}
\Pi_{+,\parallel} = N \tfrac{\widetilde{\omega}^2}{q^4} \int \frac{dk }{2\pi} \sum_{s=\pm}\left(-1 +(4p^2-q^2)\tilde I(\alpha)  \right) \Theta(\mu-k)
\end{align}
As stated earlier, the latter function is related to the density response through the Ward identity. Direct comparison with the appendix of \cite{Li2020b} shows that the Ward identity is satisfied by the above expression. One also sees that the inter and intraband polarization operators separately satisfy Eq. \eqref{relation}. In summary, one finds
\begin{gather}
\Pi^{ij}(\omega, \bm q) = \Pi_\perp(\omega, \bm q)(q^2\delta^{ij} - q^i q^j) +\Pi_\parallel(\omega, \bm q) q^i q^j \nonumber 
\end{gather}
with
\begin{gather}
q^2\Pi_{\perp}(\omega, \bm q) =  \tfrac{q^2-\omega^2}{q^2}\Pi^{00}(\omega, \bm q) + \Pi^{zz}(\omega, \bm q)\\
\Pi_\parallel(\omega, \bm q) = \frac{\omega^2}{q^4} \Pi^{00}(\omega, \bm q)
\end{gather}
as claimed in the previous section.

\subsection{Intervalley polarization operator}
Taking $J^\mu = \tau_\pm \alpha_i$ and $J^\nu = \tau_\mp \alpha_j$ with $\tau_\pm = \tfrac{1}{2}(\tau_x\pm i\tau_y)$ and performing the trace changes the sign of the terms containing factors of $\tau_z$. If we denote this intervalley polarization operator $\Pi_{II}^{ij}$ then one finds the simple result 
\begin{align}
\Pi_{II, \perp}^{ij} &=  \Pi_{\parallel}^{ij}\\
\Pi_{II, \parallel}^{ij} &= \Pi_{\perp}^{ij}
\end{align}
In other words, the transverse and longitudinal response are reversed between intra and intervalley. The same trace relations show that $\Pi^{22}=\Pi^{55}=\Pi^{66}=\Pi^{zz}$, ie the screening of the operators $\tau_\pm$ is the same as that of $\tau_z\alpha_z$.

\subsection{Solution to the RPA equations in the onsite limit}
We have the interactions
 \begin{gather}
g_1\alpha_0\otimes \alpha_0+ g_2\alpha_z\tau_z\otimes \alpha_z\tau_z +  g_4(\tau_+\otimes \tau_- + \tau_-\otimes \tau_+) \nonumber \\
g_3\left( \alpha_x\otimes \alpha_x + \alpha_y \otimes \alpha_y \right)+g_5\left( \alpha_x\otimes \alpha_x + \alpha_y \otimes \alpha_y \right) \left(\tau_+ \otimes \tau_-+\tau_- \otimes \tau_+ \right)
 \end{gather}
We consider the simple case of only onsite interactions, so only $U_{AAAA}\neq 0$; in this limit one finds $g_3=g_4=0$ for honeycomb and $g_2=g_5=0$ for kagome lattices. In each case the RPA equations decouple, so that each $g_i$ is screened separately. The result is the screened interactions:
 \begin{gather}
 \label{rpasolutionH}
 \frac{g_1}{1-g_1\Pi^{00}}\alpha_0\otimes \alpha_0+ \frac{g_2}{1-g_2\Pi^{zz}}\alpha_z\tau_z\otimes \alpha_z\tau_z \nonumber \\
  + \left(    \frac{g_5 + g_5^2 q^2 \Pi_\parallel }{1-g_5(\Pi_\parallel+\Pi_\perp)q^2 + g^2_5\Pi_\parallel\Pi_\perp q^4}\left( \alpha_x\otimes \alpha_x + \alpha_y \otimes \alpha_y \right) -  \frac{g_5^2(\Pi_\perp-\Pi_\parallel)}{1-g_5(\Pi_\parallel+\Pi_\perp)q^2 + g^2_5\Pi_\parallel\Pi_\perp q^4}(\bm q \cdot \bm \alpha) \otimes (\bm q \cdot \bm \alpha)   \right) \nonumber \\
  \times \left(\tau_+ \otimes \tau_-+\tau_- \otimes \tau_+ \right)
 \end{gather}
 for honeycomb and 
 \begin{gather}
 \label{rpasolutionK}
 \frac{g_1}{1-g_1\Pi^{00}}\alpha_0\otimes \alpha_0++\frac{g_4}{1-g_4\Pi^{zz}}(\tau_+\otimes \tau_- + \tau_-\otimes \tau_+)\nonumber \\
 + \frac{g_3 + g_3^2 q^2 \Pi_\parallel }{1-g_3(\Pi_\parallel+\Pi_\perp)q^2 + g^2_3\Pi_\parallel\Pi_\perp q^4}\left( \alpha_x\otimes \alpha_x + \alpha_y \otimes \alpha_y \right)+  \frac{g_3^2(\Pi_\perp-\Pi_\parallel)}{1-g_3(\Pi_\parallel+\Pi_\perp)q^2 + g^2_3\Pi_\parallel\Pi_\perp q^4}(\bm q \cdot \bm \alpha) \otimes (\bm q \cdot \bm \alpha) \nonumber \\
   + \left(    \frac{g_5 + g_5^2 q^2 \Pi_\parallel }{1-g_5(\Pi_\parallel+\Pi_\perp)q^2 + g^2_5\Pi_\parallel\Pi_\perp q^4}\left( \alpha_x\otimes \alpha_x + \alpha_y \otimes \alpha_y \right) -  \frac{g_5^2(\Pi_\perp-\Pi_\parallel)}{1-g_5(\Pi_\parallel+\Pi_\perp)q^2 + g^2_5\Pi_\parallel\Pi_\perp q^4}(\bm q \cdot \bm \alpha) \otimes (\bm q \cdot \bm \alpha)   \right) \nonumber \\
  \times \left(\tau_+ \otimes \tau_-+\tau_- \otimes \tau_+ \right)
 \end{gather}
 for kagome. 
 
\subsection{Cooper channel scattering amplitudes}

\subsubsection{On-site limit for honeycomb systems}
 For honeycomb systems, the $\ell$-wave scattering amplitude is 
\begin{gather}
\Gamma^\ell_{\tau_1\tau_2\tau_3\tau_4}(\omega, k,p) = \frac{1}{4}\int \frac{d\theta }{2\pi } e^{-i\ell\theta}\left( (1+e^{ i\tau_1\theta})^2\frac{g_0}{1-g_0\Pi^{00}(\omega, \bm q)}  + (1-e^{i\tau_1\theta})^2\frac{g_2}{1-g_2\Pi^{zz}(\omega, \bm q)}  \right)\delta_{\tau_1,\tau_2}\delta_{\tau_3,\tau_4}\delta_{\tau_1,\tau_3} \nonumber \\
+2e^{-i\ell\theta}\left( (1+\cos\theta)\frac{g_0}{1-g_0\Pi^{00}(\omega, \bm q)}  + (1-\cos\theta)\frac{g_2}{1-g_2\Pi^{zz}(\omega, \bm q)}  \right)\delta_{\tau_1,\tau_2}\delta_{\tau_3,\tau_4}\delta_{\tau_1,-\tau_3} \nonumber \\
+\ \ 2e^{-i\ell\theta}\left( \frac{g_5 + g_5^2 q^2 \Pi_\parallel }{1-g_5(\Pi_\parallel+\Pi_\perp)q^2 + g^2_5\Pi_\parallel\Pi_\perp q^4}\right)\delta_{\tau_1,-\tau_2}\delta_{\tau_3,-\tau_4}\delta_{\tau_1,-\tau_3}
\end{gather}
where $\bm q = \bm k -\bm p$, $\theta = \theta_{\bm k}-\theta_{\bm p}$.

\subsubsection{On-site limit for kagome systems}
For kagome we have
\begin{gather}
\Gamma^\ell_{\tau_1\tau_2\tau_3\tau_4}(\omega, k,p) = \frac{1}{4}\int \frac{d\theta }{2\pi } e^{-i\ell\theta}\left( (1+e^{ i\tau_1\theta})^2\frac{g_0}{1-g_0\Pi^{00}(\omega, \bm q)}  - 2e^{i\tau\theta}\frac{g_3 + g_3^2 q^2 \Pi_\parallel }{1-g_3(\Pi_\parallel+\Pi_\perp)q^2 + g^2_3\Pi_\parallel\Pi_\perp q^4} \right. \nonumber \\
\left. +\ (k-p)^2(1+e^{ i\tau_1\theta})^2\frac{g_3^2 (\Pi_\perp-\Pi_\parallel) }{1-g_3(\Pi_\parallel+\Pi_\perp)q^2 + g^2_3\Pi_\parallel\Pi_\perp q^4}\right)\delta_{\tau_1,\tau_2}\delta_{\tau_3,\tau_4}\delta_{\tau_1,\tau_3} \nonumber \\
+\ e^{-i\ell\theta}\left( (1+\cos\theta)\frac{2g_0}{1-g_0\Pi^{00}(\omega, \bm q)}  +\frac{2(g_3 + g_3^2 q^2 \Pi_\parallel ) }{1-g_3(\Pi_\parallel+\Pi_\perp)q^2 + g^2_3\Pi_\parallel\Pi_\perp q^4} \right. \nonumber \\
\left.-\ (k-p)^2(1+\cos\theta)\frac{2g_3^2 (\Pi_\perp-\Pi_\parallel) }{1-g_3(\Pi_\parallel+\Pi_\perp)q^2 + g^2_3\Pi_\parallel\Pi_\perp q^4}  \right)\delta_{\tau_1,\tau_2}\delta_{\tau_3,\tau_4}\delta_{\tau_1,-\tau_3} \nonumber \\
+ \ e^{-i\ell\theta}\left(  (1-\cos\theta)\frac{2g_4}{1-g_4\Pi^{zz}(\omega, \bm q)}  +\frac{2(g_5 + g_5^2 q^2 \Pi_\parallel) }{1-g_5(\Pi_\parallel+\Pi_\perp)q^2 + g^2_5\Pi_\parallel\Pi_\perp q^4}\right) \delta_{\tau_1,-\tau_2}\delta_{\tau_3,-\tau_4}\delta_{\tau_1,-\tau_3}
\end{gather}

\subsection{Eliashberg equations}
 
Neglecting self-energy corrections, the linearized Gor'kov-Eliashberg frequency dependent gap equation is given
\begin{align}
\Delta^{\ell}(i\omega_n,  k) = -T\sum_{m} \sum_p K^\ell(i\omega_n, k ; i\omega_m,   p)  \Delta^{\ell}(i\omega_m,p)
\end{align}
with the Eliashberg kernel,
\begin{align}
K^\ell(i\omega_n, k ; i\omega_m,   p) = \frac{1}{\omega_m^2 + \varepsilon_p^2} \int \frac{d\theta}{2\pi}\ e^{i\ell \theta} \ \lambda(i\omega_n - i\omega_m, k, p, \theta) 
\end{align}
where $\omega_n = \pi(2n+1)T$ are the fermionic Matsubara frequencies  \cite{Schrieffer1964,eliash, debanjan}. The frequency dependent coupling $\lambda(\omega, q)$ is defined analogously to the static coupling in the main text: $\lambda(\omega,q) = \Gamma_{\tau\tau\tau\tau}(\omega,q)$ for intravalley pairing or $\lambda(\omega,q) = \Gamma_{+-+-}(\omega,q) \pm \Gamma_{+--+}(\omega,q)$ for intervalley pairing. The functions $\Delta$, $K$ and $\Gamma$ are tensors in spin and valley space -- spin and valley indices are left implicit. The critical temperature can found by finding the value of $T$ for which the above linear equation for $\Delta$ has an eigenvalue of $-1/T$. The zero temperature gap can also be found from solving
\begin{gather}
\Delta^\ell(\omega') = -\nu_0\int \frac{\Delta^\ell(\omega)\lambda^\ell(\omega-\omega')}{\sqrt{\omega^2 - |\Delta^\ell(\omega)|^2}}  \frac{d\omega}{4\pi}
\end{gather}
where $\lambda^\ell(\omega-\omega')$ is the $\ell$-wave coupling evaluated onshell, $\varepsilon_k=\omega, \varepsilon_p=\omega'$. These equations can be derived through general assumptions about the analytic behavior of the scattering amplitude $\Gamma$ (see e.g. \cite{Schrieffer1964}). The requirements of unitarity and causality imply that as a function of frequency the amplitude must be analytic in the upper half plane, and that the amplitude has branch cuts corresponding to the threshold of particle-hole production. While ${\Gamma}$ possesses singularities as a function of $\omega'$,  these occur either for $\omega'\gg v_F|\bm{k}-\bm{p}|$, which provides a negligible contribution, or close to the edge of the particle hole continuum $\omega'\approx v_F|\bm{k}-\bm{p}|$, which is significant for a range of scattering angles $\theta_{\bm{k}}-\theta_{\bm{p}}\approx 0$ which only becomes significant when $p\approx k_F$. 

As is shown by examination of the formulae in the previous subsection, the scattering amplitudes have a roughly step-like behavior. Below, we plot the $p$-wave inter and intra valley on-shell couplings -- $\lambda^{^{\ell=1}}_{intra} = \Gamma^{\ell=1}_{\tau\tau\tau\tau}(|k|-|p|,|\bm{k}-\bm{p}|)$ as well as $\lambda^{\ell=1}_{inter} = \Gamma_{+-+-}(|k|-|p|,|\bm{k}-\bm{p}|) - \Gamma_{+--+}(|k|-|p|,|\bm{k}-\bm{p}|)$ -- for kagome and honeycomb to illustrate this phenomenon. Treating the scattering amplitude as a step function in frequency space, the frequency dependent gap equation can therefore be straightforwardly solved as in the Anderson-Morel treatment of the electron-phonon problem \cite{Morel1962}, the result being a renormalized coupling appearing in the exponential form of $T_c$ and an effective frequency cutoff, c.f. the treatment in S3 of \cite{Li2020b}.

\begin{figure}[t]

		\hspace{-0.0\textwidth}\includegraphics[height = 0.25\textwidth]{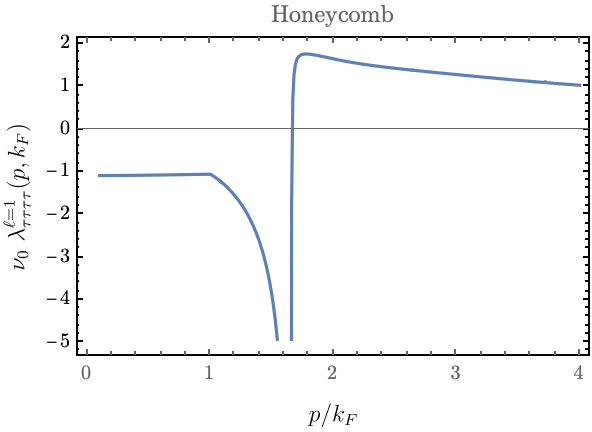} \hspace{-0.0\textwidth}\includegraphics[height = 0.25\textwidth]{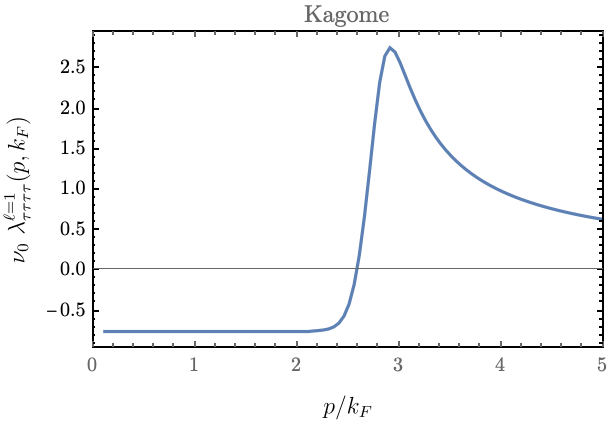} 
		
				{\hspace{-0.0001\textwidth}\includegraphics[width = 0.35\textwidth]{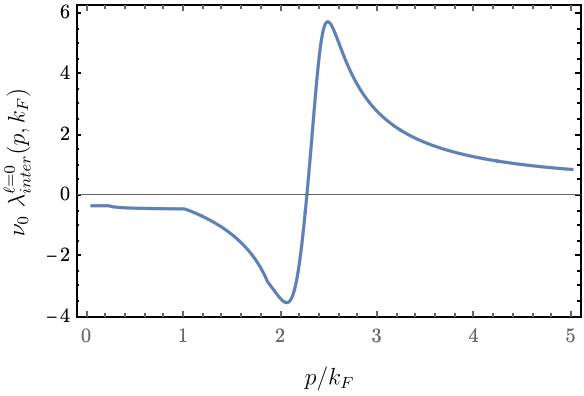}} \hspace{-0.001\textwidth}\includegraphics[width = 0.35\textwidth]{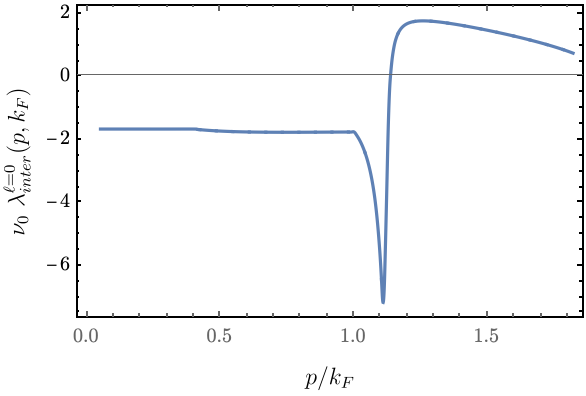} 
				
	\caption{The $p$-wave intravalley coupling function $\lambda_{\tau\tau\tau\tau}^{\ell=1}(p,k=k_F)$, and $s$-wave intervalley coupling function  $\lambda_{inter}^{\ell=0}(k,p)$ for honeycomb (left column) and kagome (right column) systems in the onsite limit. We use fine structure constant $2\pi\nu_0 e^2/\varepsilon_r=0.3$, bare couplings $g_2=0.6, g_3=1.4,g_4=0.8,g_5=1.4$, and for the intrarvalley coupling functions use $g_1=0.4$ and for the intervalley coupling functions $g_1=0.2$. The scattering amplitudes have been regularized and smoothed to avoid singularities which are inessential to the solution of the gap equation, as discussed in text.}
	\label{fig:Pi0}
\end{figure}

\newpage
\section{Real space form of the $p+i\tau p$ gap function}
\label{derivDr}

In this appendix we derive the real space form of the $p+i\tau p$ gap function used for exact diagonalization in Section \ref{Sec4}. Starting with the momentum space mean field Hamiltonian, we have
\begin{align}
H_\text{BdG}&=\sum_{\bm{k}}\varepsilon_{\bm k} f^\dag_{\bm k,\tau,s}f_{\bm k,\tau,s} + \frac{1}{2}\sum_{k < k_F,\tau,s,s'}{\Delta_k e^{i\tau \phi} e^{-i\tau \theta_{\bm{k}}} (i s_y d^\mu s_\mu)_{ss'}f^\dag_{\bm{k},\tau,s} f^\dag_{-\bm{k},\tau,s'}} +\text{h.c.}
\end{align}
where $f^\dag_{\bm k, \tau, s}$ creates an electron in the upper band, $\Delta_k$ is an overall factor depending only on the magnitude $k$, $\phi$ is the phase of the pair density wave, and $(d^x,d^y,d^z)$ is a constant 3D vector with unit length. The normal dispersioon $\varepsilon_{\bm{k}}$ is spin independent, which allows us to perform a spin rotation so that $(d^x,d^y,d^z) = -i(0,1,0)$, and the Hamiltonian decouples into two spin-diagonal terms,
\begin{gather}
H_{BdG} = H_\uparrow + H_\downarrow = \sum_{\bm{k},s}\varepsilon_{\bm k} f^\dag_{\bm k,\tau,s}f_{\bm k,\tau,s} + \frac{1}{2}\sum_{k < k_F,\tau,s,s'}{\Delta_k e^{i\tau\phi} e^{-i\tau \theta_{\bm{k}}} f^\dag_{\bm{k},\tau,s} f^\dag_{-\bm{k},\tau,s}} +\text{h.c.}
\end{gather}
Since the Hamiltonian is a sum of independent spin blocks, we discard the spin index henceforth. We now convert this expression to real space, in which the Hamiltonian becomes
\begin{gather}
H = \sum_{\bm{r},\bm{r}'}\mathcal{H}(\bm{r},\bm{r}') c^\dag_{\bm{r}} c^\dag_{\bm{r}'} + \frac{1}{2} \Delta(\bm{r},\bm{r}') c^\dag_{\bm{r}} c^\dag_{\bm{r}'} + \text{h.c.} 
\end{gather}
where $c^\dag_{\bm{r},s}$ is the full electron creation operator (as compared to $f^\dag_{\bm{k}}$ which creates an electron with momentum $\bm{k}$ in the upper band).  We assume that the only hopping term present connects nearest neighbors, i.e.  that $\mathcal{H}(\bm{r},\bm{r}')=-t$ if $\bm{r}$ and $\bm{r}'$ are nearest neighbors and zero else. It remains then to calculate the function $\Delta(\bm{r},\bm{r}')$. In momentum space, the pairing term for each spin block is given by
\begin{gather}
H_\Delta = \frac{1}{2}\sum_{k < k_F,\tau}{\Delta_k e^{i\tau \phi} e^{-i\tau \theta_{\bm{k}}} f^\dag_{\bm{k},\tau} f^\dag_{-\bm{k},\tau}}
\end{gather}
We relate the creation operator for the upper band to the full electron creation operator by
\begin{gather}
f^\dag_{\bm{k},\tau,s} = \sum_{\bm{r}}{u_{\bm{k},\tau}(\bm{r}) c^\dag_{\bm{r},s}} 
\end{gather}
where $u_{\bm{k},\tau}(\bm{r}) $ is the wavefunction for the upper Dirac band.  Substituting this relation into the pairing term,
\begin{gather}
H_\Delta = \frac{1}{2} \sum_{k,\tau,\bm{r},\bm{r}'}{ \Delta_k e^{-i\tau\phi}e^{-i\tau\theta_{\bm{k}}} u_{\bm{k},\tau}(\bm{r}) u_{-\bm{k},\tau}(\bm{r}') c^\dag_{\bm{r},s} c^\dag_{\bm{r}',s}} 
\end{gather}
from which it follows immediately that
\begin{gather}
\Delta(\bm{r},\bm{r}') = \sum_{k,\tau}{\Delta_k
e^{-i\tau\phi} e^{-i\tau\theta_{\bm{k}}} u_{\bm{k},\tau}(\bm{r}) u_{-\bm{k},\tau}(\bm{r}')} \ \ .
\end{gather}
The upper band eigenstates of the Dirac Hamiltonian $\mathcal{H} = v\tau \bm{k}\cdot\bm{\alpha}$ are given in the coordinate representation in terms of the wavefunctions $e^{i\tau\bm{K}_1\cdot\bm{r}} \varphi_{\tau \alpha}(\bm{r})$ by
\begin{align}
u_{\bm{k},+}(\bm{r}) &= \tfrac{1}{\sqrt{2}} \left(\varphi_{++}(\bm{r}) + e^{i\theta_{\bm{k}}} \varphi_{+-}(\bm{r})\right) e^{i(\bm{k}+\bm{K})\cdot\bm{r}} \ \ , \nonumber \\
u_{\bm{k},-}(\bm{r}) = u^*_{-\bm{k},+}(\bm{r}) &=\tfrac{1}{\sqrt{2}} \left(\varphi^*_{++}(\bm{r}) - e^{-i\theta_{\bm{k}}} \varphi^*_{+-}(\bm{r})\right) e^{i(\bm{k}-\bm{K})\cdot\bm{r}}
\end{align}
where $\varphi_{\tau\alpha}(\bm{r})$ are periodic functions under lattice translations. Thus
\begin{gather}
\Delta(\bm{r},\bm{r}') = \sum_{\bm{k}}{}\Delta_k\{
\tfrac{1}{2}e^{i\{\bm{K}\cdot(\bm{r}+\bm{r}') + \bm{k}\cdot(\bm{r}-\bm{r}') + \phi -\theta_{\bm{k}}\}}
\left[
(\varphi_{++}(\bm{r}) + e^{i\theta_{\bm{k}}}\varphi_{+-}(\bm{r}))
(\varphi_{++}(\bm{r}') - e^{i\theta_{\bm{k}}} \varphi_{+-}(\bm{r}'))\right] \nonumber \\
+ 
\tfrac{1}{2}e^{i\{ -\bm{K}\cdot(\bm{r}+\bm{r}')+\bm{k}\cdot(\bm{r}-\bm{r}')  -\phi +\theta_{\bm{k}}\}}
\left[
(\varphi^*_{++}(\bm{r}) - e^{-i\theta_{\bm{k}}} \varphi^*_{+-}(\bm{r}))
(\varphi^*_{++}(\bm{r}') + e^{-i\theta_{\bm{k}}} \varphi^*_{+-}(\bm{r}'))\right]\}
\end{gather}
Performing the summation over $\bm{k}$ and introducing the functions
\begin{gather}
\int{\Delta_k e^{i\bm{k}\cdot\bm{r}} \frac{d^2\bm{k}}{(2\pi)^2}} = \frac{1}{2\pi}\int{\Delta_k J_0(kr) kdk} = \Phi_0(r) \ \ , \nonumber \\ 
i^\mp\int{\Delta_k e^{i\left[\bm{k}\cdot\bm{r} \pm \theta\right]} \frac{d^2\bm{k}}{(2\pi)^2}} = \frac{1}{2\pi}\int{\Delta_k J_1(kr) kdk} = \Phi_1(r)
\end{gather}
we find
\begin{gather}
\Delta(\bm{r},\bm{r}') \nonumber \\
= 
\tfrac{1}{2}e^{i\{\bm{K}\cdot(\bm{r}+\bm{r}') + \phi \}}\left[
(\varphi_{++}(\bm{r}) \varphi_{++}(\bm{r}') - \varphi_{+-}(\bm{r}) \varphi_{+-}(\bm{r}')) i\Phi_1(r)
+ (\varphi_{+-}(\bm{r}) \varphi_{++}(\bm{r}') - \varphi_{++}(\bm{r}) \varphi_{+-}(\bm{r}'))\Phi_0(r)\right] \nonumber \\
+ 
\ \tfrac{1}{2} e^{-i\{\bm{K}\cdot(\bm{r}+\bm{r}')+ \phi\}}\left[
-(\varphi_{++}(\bm{r}) \varphi^*_{++}(\bm{r}') - \varphi^*_{+-}(\bm{r}) \varphi^*_{+-}(\bm{r}'))
i\Phi_1(r)
+ (\varphi^*_{++}(\bm{r}) \varphi^*_{+-}(\bm{r}')
- \varphi^*_{+-}(\bm{r}) \varphi^*_{++}(\bm{r}') ) \Phi_0(r) \right] \nonumber \\
=\text{Re}\{
e^{i\{\bm{K}\cdot(\bm{r}+\bm{r}')+ \phi+\frac{\pi}{2}\}}\left[
(\varphi_{++}(\bm{r}) \varphi_{+-}(\bm{r}') - \varphi_{+-}(\bm{r}) \varphi_{+-}(\bm{r}')) |\Phi_1(r)|
+ (\varphi_{++}(\bm{r})\varphi_{+-}(\bm{r}') - \varphi_{+-}(\bm{r}) \varphi_{++}(\bm{r}'))|\Phi_0(r)|\right]\} 
\end{gather}
The function $|\Phi_0(r)|$ is maximum for $r = 0$ and goes to zero over length scales $r \sim \pi/k_F$ while $|\Phi_1(r)|$ is small for small $r$ and increases to a maximum at $r \sim\pi/k_F$. To simplify $\Delta(\bm{r},\bm{r}')$ for the purposes of exact diagonalization, we will neglect the pairing correlations at large separations, discarding the term containing $\Phi_1$ and set $\Phi_0 \rightarrow \Delta'$ with $\Delta'$ being some average value. With these simplifications we finally arrive at the real space Hamiltonian
\begin{gather}
H = \sum_{\langle \bm{r},\bm{r}'\rangle}{
-t c^\dag_{\bm{r}} c_{\bm{r}'} + \frac{1}{2}\Delta(\bm{r},\bm{r}') c^\dag_{\bm{r}} c^\dag_{\bm{r}'} + \text{h.c.} } \ \ ,\nonumber \\
\Delta(\bm{r},\bm{r}') =\Delta' \text{Re} \{e^{i\{\bm{K}\cdot(\bm{r}+\bm{r}') + \phi+\frac{\pi}{2}\}} \left[\varphi_{++}(\bm{r}) \varphi_{+-}(\bm{r}') -\varphi_{+-}(\bm{r})\varphi_{++}(\bm{r}')\right] \} \ \ .
\label{final_real}
\end{gather}
Note that $\Delta(\bm{r},\bm{r}') = -\Delta(\bm{r}',\bm{r}) = \Delta^*(\bm{r},\bm{r}')$, and with $t$ real, the Hamiltonian is invariant under complex conjugation, conforming with our claim in the main text that this superconducting phase lies in class BDI.

Below we present $\Delta(\bm{r},\bm{r}')$ explicitly for the honeycomb and Kagome lattices. The procedure is  simply to calculate the eigenfunctions $\varphi_{\tau \alpha}(\bm{r})$ of the normal state Hamiltonians at the Dirac points, and insert into Eq. \eqref{final_real}.

\subsection{Honeycomb lattice}
For the honeycomb lattice, we have
\begin{gather}
\varphi_{+ +}(\bm{r}) =\begin{cases}
1 \ \  \ \bm{r} \in A\\
0 \ \  \ \bm{r} \in B
\end{cases} \  \ \
\varphi_{+-}(\bm{r}) =\begin{cases}
0 \ \  \ \bm{r} \in A\\
1 \ \  \ \bm{r} \in B
\end{cases} 
\end{gather}
For $\bm{r}\in A,\bm{r}'\in B$ we therefore get
\begin{gather}
\Delta(\bm{r},\bm{r}') = \text{Re} \{ e^{i\{\bm{K}\cdot(\bm{r}+\bm{r}') + \phi+\frac{\pi}{2}\}}\} = \sin(\bm{K}\cdot(\bm{r}+\bm{r}') + \phi)
\end{gather}
and hence
\begin{gather}
H = \sum_{\langle \bm{r},\bm{r}'\rangle}{
-t c^\dag_{\bm{r}} c^\dag_{\bm{r}'} + \Delta'\left[\sin(\bm{K}\cdot(\bm{r}+\bm{r}') + \phi) c^\dag_{\bm{r}} c^\dag_{\bm{r}'} + \text{H.c.}\right]}
\end{gather}
as per Eq. \eqref{realDH}.

\subsection{Kagome lattice}

We have
\begin{align}
\varphi_{\tau \alpha}(\bm{r}) = \begin{cases}
1 \ \  &\ \bm{r} \in A \\
-e^{-\frac{2\pi i \alpha}{3}}\ \  &\  \bm{r} \in B \\
-e^{\frac{2\pi i \alpha}{3}}\ \  &\ \bm{r}\in C
\end{cases}
\end{align}
so the real space gap function is
\begin{gather}
\Delta(\bm{r},\bm{r}') = \begin{cases}
\cos(\bm{K}\cdot(\bm{r}+\bm{r}') + \phi ) \ \ , \ \bm{r}\in A, \bm{r}'\in B \\
\cos(\bm{K}\cdot(\bm{r}+\bm{r}') +\phi+\pi) \ \ , \ \bm{r}\in B, \bm{r}'\in C \\
\cos(\bm{K}\cdot(\bm{r}+\bm{r}') + \phi ) \ \ , \ \bm{r}\in C, \bm{r}'\in A
\end{cases}
\end{gather}
and hence
\begin{gather}
H = \sum_{\langle \bm{r},\bm{r}'\rangle}{-t c^\dag_{\bm{r}} c_{\bm{r}'} 
+ \Delta'
\cos(\bm{K}\cdot(\bm{r}+\bm{r}') + \phi) c^\dag_{\bm{r}} c^\dag_{\bm{r}'} + \text{h.c.}}
\end{gather}
with the appropriate orderings of $\bm{r}$ and $\bm{r}'$ to take account of the case with a relative $\pi$ phase -- as per Eq. \eqref{realDK}


\end{document}